\documentclass{aa}
\usepackage{txfonts}
\usepackage[dvips]{graphicx}
\usepackage{natbib}
\bibpunct{(}{)}{;}{a}{}{,}
\defcitealias{SVM99}{SVM}
\defcitealias{now00}{N00}
\defcitealias{pot03}{P03}

\begin{document}
\authorrunning{M. Axelsson et al.}
\title{Evolution of the 0.01 -- 25 Hz power spectral components \\
in Cygnus X-1}  
\author{Magnus Axelsson \and Luis Borgonovo \and Stefan Larsson}
\institute{Stockholm Observatory, AlbaNova, SE- 106 91 Stockholm, Sweden}
\offprints{M. Axelsson,\\ \email{magnusa@astro.su.se}}
\date{Received 12 November 2004/ Accepted 2 April 2005}

\abstract{Analyzing the archival data from the \textit{Rossi X-ray Timing 
Explorer} 
(RXTE), we study the power density spectra (PDS) of Cygnus X-1 from 1996
to 2003 in the frequency range of 0.01 -- 25 Hz. Using a model consisting 
of one or two Lorentzians and/or an exponentially cut-off power-law, we are 
able to achieve a good fit to the PDS during the observations. With our 
model we are also able to track the evolution of the Lorentzian components 
through all spectral states of the source. We confirm
the relation between characteristic frequencies seen both in black hole
candidate and neutron star sources, and show the changes in
this relation during the transitional and soft states of the source. The 
connection between the Lorentzian components is investigated by analyzing 
similarities and differences in their behavior. We find that the
spectral state of the source can be uniquely determined from the parameters
of the these components. The parameter correlations can all be described
by continuous functions, which differ between components. We discuss our 
results in the context of relativistic precession model for the accretion 
disk, and show a remarkable agreement between the model prediction and the 
data in the hard state. We estimate a value for the specific angular momentum 
of $a_*=0.49$ ($-0.57$) in the case of prograde (retrograde) rotation 
and an estimate for the inner radius of 22 to 50 (25 to 55) gravitational
radii. Additional assumptions are required to explain the soft 
state data, and attempting to invoke rotational reversal for state 
transitions shows that it is insufficient to explain the differences between 
the hard and soft state data.
}

\maketitle

\section{Introduction}

In many X-ray binary systems, the existence of distinct spectral states has 
been well documented. Often quoted as a prototype black hole system, 
\object{Cygnus X-1} has been extensively observed and modeled. The source 
exhibits two main spectral states, usually referred to as hard and soft, 
with a brief intermediate state \citep[sometimes interpreted also as a very 
high state, see e.g. discussion in][ and references therein]{zg04} during the 
transitions. Several models have been proposed to explain the spectral states 
and transitions, with the two main components being a geometrically thin, 
optically thick accretion disk and a hot inner flow or corona. Seed photons 
from the disk are believed to gain energy in the corona through Comptonization 
\citep{sha76}. The
models vary in the geometry and properties of mainly the corona/comptonizing 
region \citep[for a review on coronal models, especially in regard to 
timing characteristics, see][]{pou01}. However, changes in the inner radius
of the accretion disk are a common source of 
variability in many models \citep[e.g.][]{pou97,esi98,chu01,zdz02}.       

The changes in spectral state are evident also in the power density 
spectrum (PDS). In the hard state the PDS is characterized by a 
flat-topped component up to $\sim 0.2$ Hz, where it breaks to a $f^{-1}$ 
slope that steepens at a few Hz \citep[e.g.][]{bh90,now99}. In the soft 
state the PDS is dominated by a $f^{-1}$ component 
that steepens around 10 Hz \citep[e.g.][]{cui97a}. The 1996 
state transition also revealed an intermediate PDS, showing a $\sim f^{-1}$ 
slope at lower frequencies, a flatter component around $0.3-3$ Hz, and 
above that similar to the hard state PDS. In many observations there is 
evidence for a quasi-periodic oscillation (QPO) at a few Hz 
\citep{bel96,cui97b,cui99}. 

Previous studies of the PDS of Cyg X-1 have shown correlations 
between temporal and spectral components \citep[e.g.][]{gil99} and also 
between different temporal features. Correlations between features in the 
hard state PDS were first reported by \citet{wvdk99}. To resolve the origin 
of the temporal features, it is vital to understand their long term behavior.
\citet[][ hereafter P03]{pot03} recently presented a large scale study of 
the hard state PDS and also included data from `failed state transitions'. 
However, no study has been presented where the PDS for all spectral states 
of Cyg~X-1 are consistently modeled. To this end, we conducted a 
systematic analysis of all the available data from the 
\textit{Rossi X-Ray Timing Explorer} (RXTE) satellite observations of 
Cyg~X-1 in the HEASARC public archives.  
Using a model consisting of a power-law and two Lorentzian 
profiles, we were able to fit the majority of PDS, and to follow the 
components from hard state through the transitions and back. This 
allowed us to expand on previous results and monitor the evolution 
of the components. 

Our aim in this paper is to provide a consistent description of the
PDS of Cyg~X-1 in all states, and through this study the
evolution of the PDS throughout the time it has been observed by RXTE.
We begin with a brief description of the available data, the
computation of lightcurves and power spectra, and a presentation
of our models in Sect.~\ref{analysis}. This is followed by a presentation 
of our results, where we show the behavior of the different model components 
(Sect.~\ref{results}). In Sect.~\ref{discussion} we follow with a discussion
of the components, as well as putting our results in the context of
physical models. Finally we summarize our findings in Sect.~\ref{conclusion}.

\section{Observations and Data Analysis}
\label{analysis}
\subsection{The archival data}
Since its launch on December 30, 1995, and until the end of April 2004, the 
RXTE satellite has observed Cyg~X-1 in more than 750 pointed observations. 
The satellite has an orbital period of about 5.8 ks, but due to its low 
circular orbit, typical on-source time
in a pointed observation is 3 -- 4 ks. The satellite also carries an
All-Sky Monitor (ASM) instrument on board, which continuously scans
the sky, covering 80\% every orbit. 

In this study we have used all public pointed observations, 
with data from 
the Proportional Counter Array \citep[PCA,][]{jah96}. The PCA consists of
five identical Proportional Counter Units (PCUs), and is sensitive
in the 2-60 keV energy range. The observations
span the time from January 8, 1996 to April 3, 2004. The data mode
used was the `Generic Binned' mode.\footnote{For one observation run in 
epoch 1, P10240, only data from the single-bit mode was available, with the 
energy range $\sim 2-9.5$ keV.} For the count (hardness) ratios 
(see Sect.~\ref{hardspec}) we used data in the Standard2 configuration. 
Due to the
different configurations and epochs for the observations, it was not
always possible to obtain an exact match between energy ranges. We found 
that choosing the energy range of $\sim 2-9$ keV allowed us to maintain the
same range fairly well throughout the observations, with any deviations 
appearing in the upper boundary. We note that the shape of the PDS of 
Cyg~X-1 is not very sensitive to the width of the energy band in this 
region \citetext{\citealt{now99}; \citetalias{pot03}}, so the
slight variations in the upper boundary should not affect our results. For 
the hardness ratios, the use of Standard2 data enabled us to keep the
same energy ranges throughout the epochs. The channels and corresponding
energy ranges are presented in Table~\ref{chantable}.

\begin{table}
\caption{Channel and energy ranges used in the count ratios and PDS.} 
\label{chantable}
\centering
\begin{tabular}{l r r r}
\hline \hline
{} & \multicolumn{2}{c}{Count ratios} & PDS\\
{} & Lower band & Upper band & Upper channel$^{\mathrm{a}}$\\
\hline
\multicolumn{4}{l}{PCA Epoch 1}\\
channel & $0-10$ & $30-60$ & 35\\
energy (keV) & $\sim2-4$ & $\sim9-20$ & 9.5\\
\hline
\multicolumn{4}{l}{PCA Epoch 2}\\
channel & $0-6$ & $25-54$ & 29\\
energy (keV) & $\sim2-4$ & $\sim9-20$ & 9.3\\
\hline
\multicolumn{4}{l}{PCA Epoch 3}\\
channel & $0-6$ & $21-50$ & $21-26$\\
energy (keV) & $\sim2-4$ & $\sim9-20$ & $8.0-9.8$\\
\hline
\multicolumn{4}{l}{PCA Epoch 4}\\
channel & $0-5$ & $16-42$ & $20-22$\\
energy (keV) & $\sim2-4$ & $\sim9-20$ & $8.8-9.7$\\
\hline
\multicolumn{4}{l}{PCA Epoch 5}\\
channel & $0-5$ & $17-44$ & $21-22$\\
energy (keV) & $\sim2-4$ & $\sim9-20$ & $9.3-9.7$\\
\hline
\end{tabular}
\begin{list}{}{}
\item[$^{\mathrm{a}}$] Only the upper channel and energy boundaries
are given, the lower boundary is $\sim2$ keV in all datasets. For
epochs 3 to 5, the upper boundary varied according with the channel
binning used in the observation, and the ranges of upper channels
and corresponding energies are given.
\end{list}
\end{table}

We used the standard RXTE analysis software \texttt{FTOOLS}, Version 5.2.
Lightcurves and X-ray spectra were extracted using standard
screening criteria: a source elevation $>10\degr$, a pointing offset
$<0\fdg01$ and a South Atlantic Anomaly exclusion time of 30 minutes. 
For the PDS, 
the lightcurves were extracted with a time resolution of 10 ms
(for a few observations a time resolution of 16 ms was the
highest possible, due to the available data mode). The lightcurves
used in determining count ratios were extracted with 16 s
time resolution. To allow comparison between observations during 
different epochs and using different numbers of PCUs, the lightcurves 
were all normalized to one PCU.

The data taken from the ASM instrument are the one-day average
count rates. All three bands have been combined, and the energy
range is $\sim 2-12$ keV. 
  
\subsection{Calculating the power density spectra}

To calculate the power density spectra the high-resolution light
curves were divided into segments of $2^{13}$ bins ($\sim 82$ s) and
a Fourier transform performed on each segment. The resulting PDS
were then gathered into groups of 10 and each group averaged into one
final PDS. Unfortunately, calculating the PDS over short segments 
results in a low frequency resolution and we were therefore not able 
to model any narrow features in the PDS. Reducing the number of averaged 
segments increases the statistical fluctuations. However, our approach 
allows changes in the PDS on timescales as short as $\sim 15$ minutes 
to be monitored. Our preliminary studies showed that such changes
do in fact occur, and as a result we accepted the lower 
signal-to-noise ratio. This is further discussed in Sect.~\ref{states}.  

The frequency range in our observations is 0.01-25 Hz. The PDS are 
corrected for the Poisson level. In principle, dead-time effects 
should be considered when determining the Poisson level.
For the frequency range used here this correction
is very small ($\la 2 \%$) compared to the statistical fluctuations, 
and we therefore disregard it. As the effect of deadtime increases with 
frequency, we set our upper limit to 25 Hz. For a discussion on
dead-time influence on PDS see \citet{vik94} and \citet{zha95}, and for 
its relevance in the case of the PCA detector \citet{zha96} and 
\citet{jer00}.    

The PDS normalization used in this paper is that of \citet{bh90} and 
\citet{miy92}. In this normalization, integrating over a frequency range
gives the square of the fractional root mean square (RMS) variability
in that frequency interval. The PDS are shown both in units of power 
versus frequency and frequency times power versus frequency 
\citep[$fP_f$ versus $f$, see e.g.][]{bel97}. The latter method of presenting
the PDS more clearly shows the contribution of variations per frequency
decade to the total RMS of the source. The PDS have been
rebinned to semi-logarithmically spaced bins.

\begin{figure*}
\centering
\includegraphics[width=17cm]{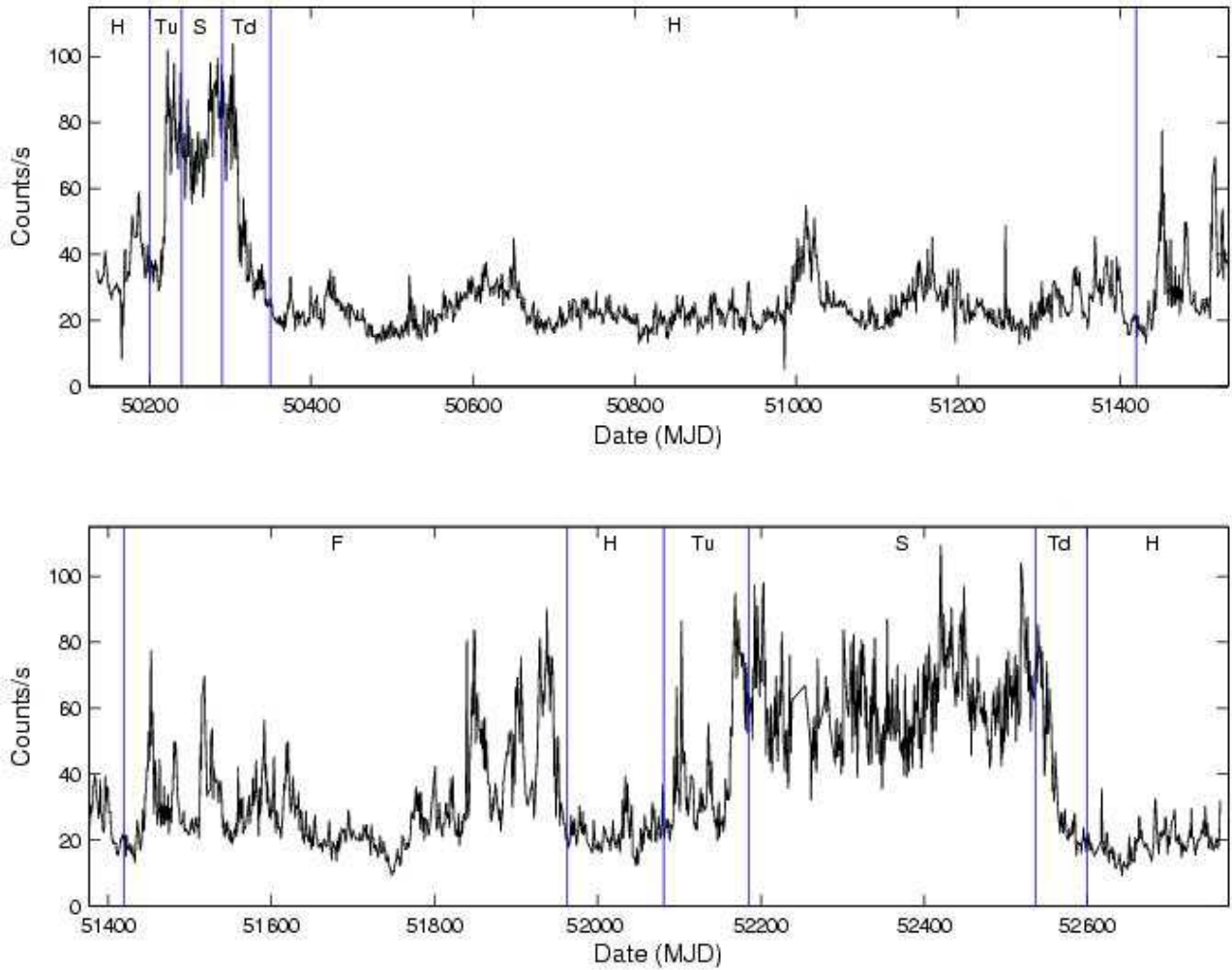}
\caption{ASM lightcurve for Cyg~X-1 from MJD 50\,135 
(Feb 22, 1996) to MJD 52\,763 (May 4, 2003). The different 
groups are marked in the plot, with $H$ signifying hard state, $Tu$ and 
$Td$ transitions from hard to soft and soft to hard respectively, $S$ the 
soft state and $F$ signifies a period of large flares in the hard state.} 
\label{asmlc}
\end{figure*}

The observations were provisionally divided into groups, roughly depending 
on the state of the source. Figure~\ref{asmlc} shows the ASM lightcurve 
of Cyg~X-1, and the vertical lines show boundaries between the different 
groups. The intervals around the transitions are deliberately chosen 
wide in order to encompass hard and soft state PDS before and after 
each transition. A period of 
intense flaring activity in the hard state \citep[and possibly one or
more brief transitions to the soft state during MJD 51\,830 -- MJD 51\,960, 
see][]{cui02} was treated separately. The reduction resulted in a total of 
2016 PDS, of which 1324 were from the hard state, 226 in flares, 191 during 
transitions, and 275 in the soft state. Approximately 2\% of the computed
PDS were of too low quality to be used in the analysis. 
 
\subsection{Modeling}
\label{models}
The PDS of Cyg~X-1 have traditionally been described using a 
double-broken power-law in the hard state \citep[e.g.][]{bh90,now99} and
an $f^{-1}$ slope broken at higher frequencies in the soft state 
\citep{cui97a,cui99,chu01}. Building on results from comparisons between 
neutron star (NS) and black hole candidate (BHC) sources 
\citep[][~and references therein]{pbk99,wvdk99} it has been shown
that models using Lorentzian profiles provide a good description
of the hard state of Cyg~X-1 \citetext{\citealt{now00}, hereafter 
N00; \citetalias{pot03}}.
However, these models do not fit the soft state of Cyg~X-1, and 
during large flares and state transitions an additional power-law 
component is required \citepalias{pot03}. 

While there is evidence for several Lorentzian profiles in the PDS of
Cyg~X-1, generally only two dominate at a given time 
\citetext{\citetalias{now00}; \citealt{gil99}}. As a base for our model 
we therefore consider the sum of two Lorentzian profiles of the form
\begin{equation}
L_i(f)=\frac{H_iW_i\nu_i}{(f-\nu_i)^2+fW_i\nu_i} 
\end{equation}
\noindent where $H_i$ is the value of $fP_f$ at the peak frequency $\nu_i$. 
In the $fP_f$ representation, the dimensionless parameter $W_i$ gives 
a measure of the width of the profile. The parametrization is chosen 
such that for a given $W_i$ and $H_i$, the integral of the Lorentzian (and 
thus its fractional contribution
to the total RMS variability) is the same for any value of $\nu_i$.
Note that while this parametrization differs from others previously used
\citepalias[e.g.][]{now00,pot03}, it has the advantage of having a direct 
interpretation of the parameters in the $fP_f$ representation.

We have chosen to use the peak frequency, which is the frequency where the 
contribution of the Lorentzian to the total RMS is at its maximum, rather 
than the centroid frequency, since this seems to be the important parameter
in terms of frequency relations \citetext{\citetalias{now00}; \citealt{vst02}}. 
In the case of narrow features, the peak and centroid frequencies
will be virtually identical. For broad features (such as breaks in the 
standard PDS representation), it is the peak frequency which best represents 
the characteristic frequency of the feature. In such cases the centroid 
frequency has no direct physical meaning, and will approach zero. For further
discussion on the peak and centroid frequencies see \citet{bel02}.
  
The fitting is done in the $fP_f$ parameter space, where
our parameters have the interpretation described above. Conversion
to `standard' Lorentzian parameters shows the advantage of using
the peak frequency. When the width parameter $W_i$ is greater than 2, the 
centroid frequency for the Lorentzian in the PDS space becomes negative. 
Although not a problem mathematically, this underscores the fact that in 
the context of broad components it is the peak frequency that has a 
physical interpretation. In principle we could have frozen the centroid 
frequency to zero in such cases, but we left it free in order to better 
follow the variations of the PDS at very low frequencies. 
 
In addition to the Lorentzian profiles, when necessary (large flares,
transitions and in the soft state) we have included a cut-off power-law 
component of the form
\begin{equation}
Pl(f)=Af^{-\alpha} e^{-f/f_c}
\end{equation}
\noindent where A is the normalization constant, $\alpha$ is the 
power-law index, and $f_c$ is the turnover frequency. Studies at 
longer timescales of Cyg~X-1 using RXTE/ASM data \citep{rei02} show 
that at low frequencies (below $\sim 10^{-5}$ Hz) the PDS is approximately 
described by a $\sim f^{-1}$ 
power-law, and there is evidence that the PDS follows this index to 
higher frequencies during transitions and in the soft state \citep{zdzprep}.
A simple $f^{-1}$ power-law does not fit the soft state PDS above 
$\sim 10$ Hz, and we therefore introduce an exponential cut-off.     

\begin{figure}
\resizebox{\hsize}{!}{\includegraphics{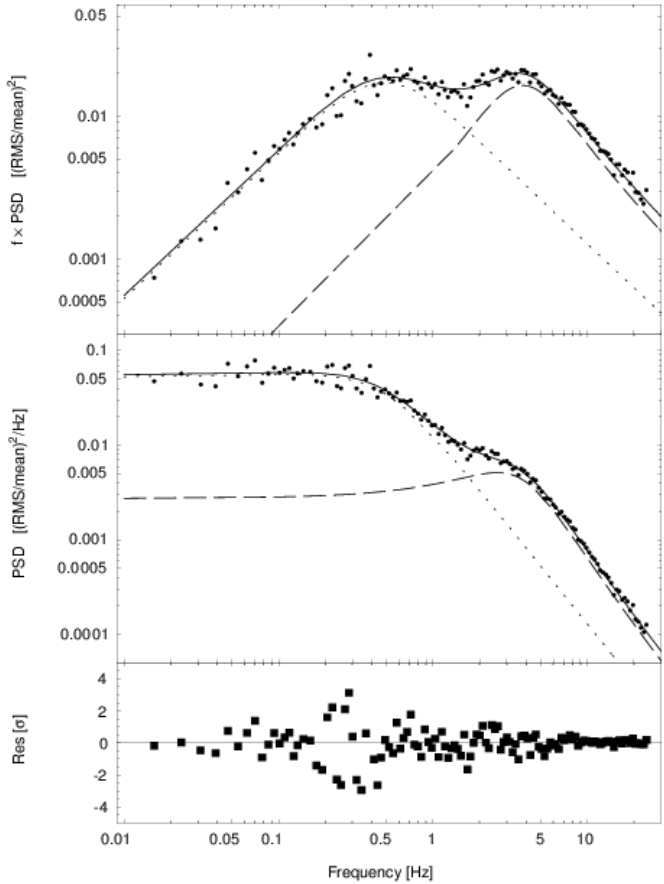}}
\caption{Example of fit to hard state PDS using two Lorentzian profiles 
(model 1). This is the most frequent state of the PDS, and the majority of 
our observations are fit with this model. Hardness ratios (9 -- 20 keV/
2 -- 4 keV) for this model fall in the range of 0.50 -- 1.1. The example is 
from an observation made on MJD 50\,180.}
\label{model2ex}
\end{figure}

\begin{figure}
\resizebox{\hsize}{!}{\includegraphics{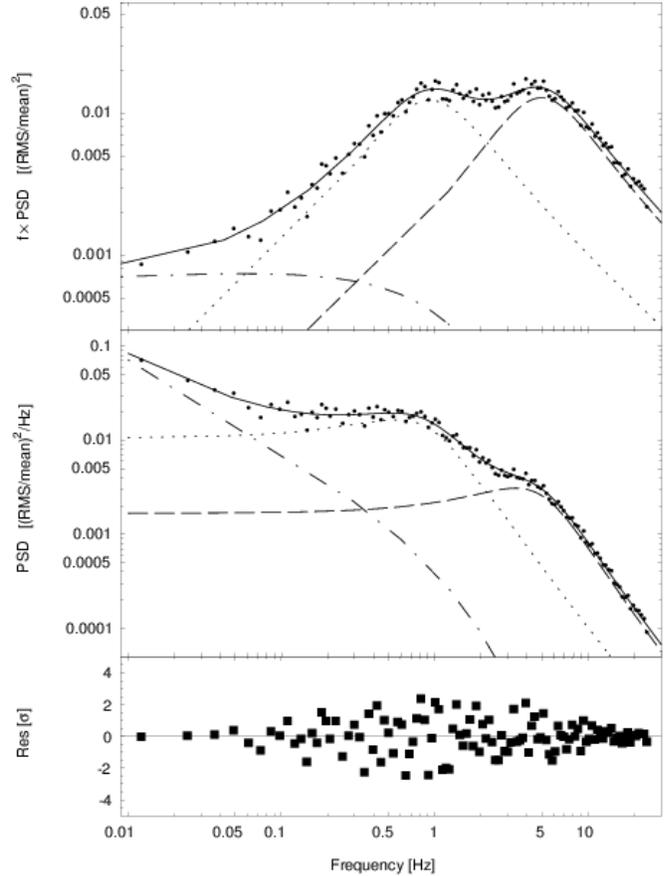}}
\caption{Example of fit to a transition PDS using a cut-off power-law
and two Lorentzian profiles (model 2). The PDS of Cyg~X-1 take on this 
appearance during state transitions and larger flares in the hard state
(failed transitions). Hardness ratios for these fittings fall in the 
range of 0.25 -- 0.45. The example is from an observation made on 
MJD 50\,226.}
\label{model3ex}
\end{figure}

\begin{figure}
\resizebox{\hsize}{!}{\includegraphics{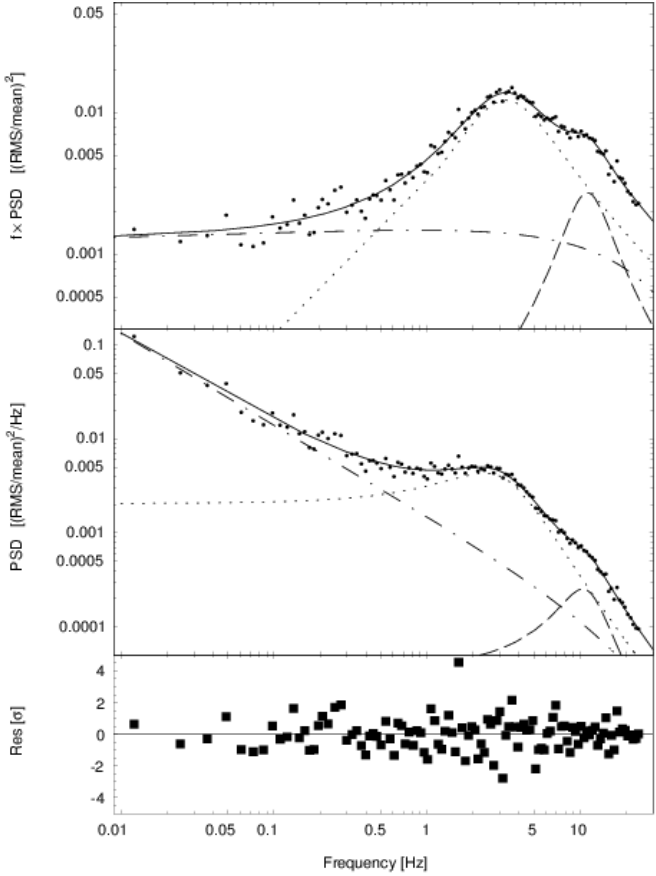}}
\caption{Example of fit to soft state PDS using a cut-off power-law
and two Lorentzian profiles (model 2). Almost 20 \% of the soft state 
observations also allowed a fit with this model. The Lorentzian 
components have shifted compared to Fig.~\ref{model3ex}, and the power-law 
component extends to higher frequencies than during transitions. Hardness 
ratios for these fittings fall within the range of 0.20 -- 0.30. The 
example is from an observation made on MJD 52\,341.}
\label{model3bex}
\end{figure}

\begin{figure}
\resizebox{\hsize}{!}{\includegraphics{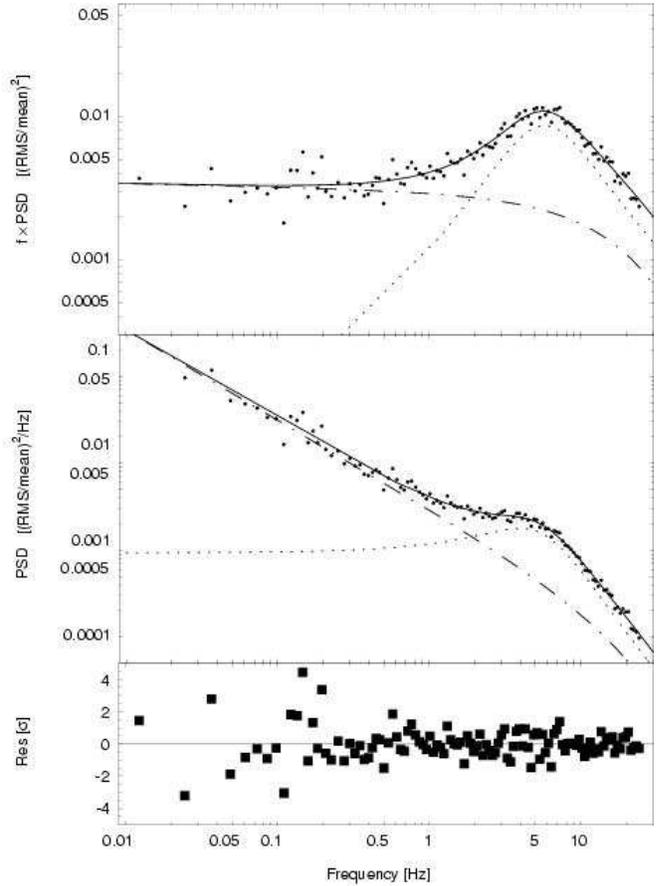}}
\caption{Example of fit to soft state PDS using a cut-off power-law and 
one Lorentzian profile (model 3). Almost half the soft state observations
allowed a fit using this model. Hardness values for this model 
fall in the range of 0.14 -- 0.26. The example is from an observation 
made on MJD 52\,341.}
\label{model4ex}
\end{figure}

\begin{figure}
\resizebox{\hsize}{!}{\includegraphics{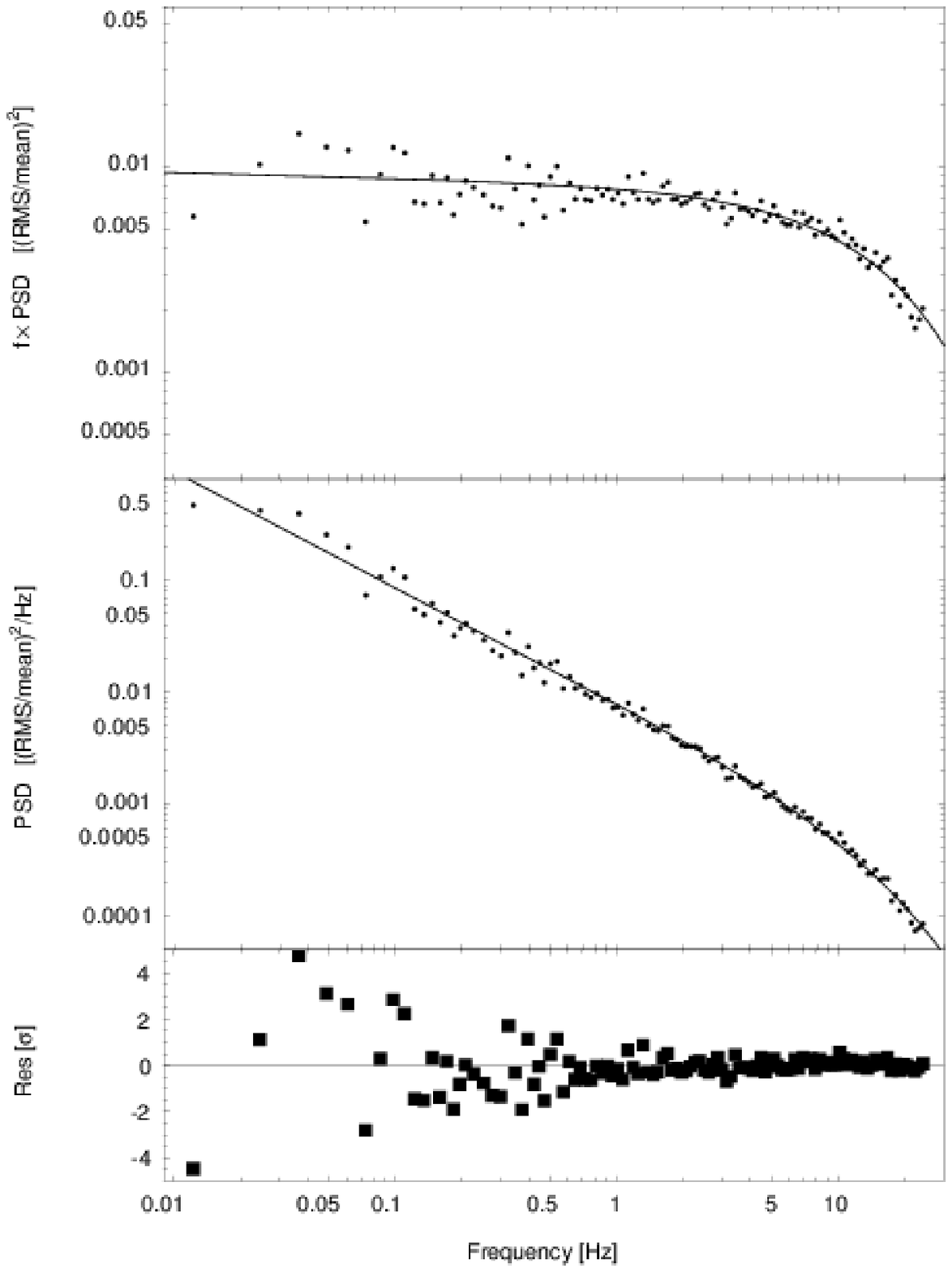}}
\caption{Example of fit to soft state PDS using only a cut-off power-law 
(model 4). This is the canonical soft state PDS appearance for Cyg~X-1, 
however only 35\% of the observations made during the soft state allowed 
a fit using this model. Hardness values for this model fall in the range 
of 0.11 -- 0.23. The example is from an observation made on MJD 52\,257.}
\label{model5ex}
\end{figure}

As will be described in Sect.~\ref{results}, when the second Lorentzian 
shifts to higher frequencies in the soft state, its contribution diminishes 
and it becomes difficult to constrain. Some PDS were therefore fit with 
only one Lorentzian profile plus a power-law component. In addition, 
the canonical soft state PDS exhibiting only a $f^{-1}$ slope do not 
require any Lorentzian profile. Table~\ref{components} lists the component 
combinations used in each state, and example fits for each of the four 
models are shown in Figs.~\ref{model2ex} to~\ref{model5ex}. 

\begin{table}
\caption{Combinations of components used.} 
\label{components}
\centering
\begin{tabular}{l l l}
\hline \hline
Model & Components & States used\\
\hline
1 & 2 Lorentzians & hard\\
2 & Power-law+2 Lorentzians & all\\
3 & Power-law+Lorentzian & soft\\
4 & Power-law & soft\\
\hline
\end{tabular}
\end{table}


While it is clear that Cyg~X-1 exhibits several spectral states, there
is no unique classification between them. Adding to the difficulty are 
the large flares or 'failed state transitions', where the source
appears to begin a transition but not complete it. We have chosen to
call these episodes flares, and reserve the use of transition for
cases when the source changes state. In this study we have attempted to 
use the PDS when determining the state of the source. For the canonical 
hard and soft states, there is no difficulty as their PDS are clearly 
separate and appear only in their respective state. The situation during 
the transitions (and flares) is by its very nature more ambiguous.
In the following, we distinguish between hard and transitional state
PDS by the power-law component appearing only in the latter. The 
same criterion is used to distinguish flares, with the difference
between flares and transitions being that the source does not change
state after a flare. The distinction between transitional and soft 
states is not as clear-cut. We have in this study assigned the PDS where 
the second Lorentzian component is not seen or significantly weaker than 
the first one to the soft state, along with the canonical PDS consisting
only of the power-law component. In our classification, Fig.~\ref{model2ex} 
is a hard state PDS, Fig.~\ref{model3ex} a PDS from a flare or transition, 
and Figs.~\ref{model3bex} to \ref{model5ex} soft state PDS. We shall compare
this classification to others made from a spectral standpoint in 
Sect.~\ref{pdsevol}.

\subsection{Hardness and Spectral Index}
\label{hardspec}

The energy spectrum of Cyg~X-1 in the hard state can be represented as
a soft thermal component and a hard power-law component with index
$\Gamma \sim 1.6$ and a high-energy cutoff. There is evidence of an
iron fluorescence line at 6.4 keV and a broad Compton reflection
component around 30 keV \citep[e.g.][]{bal95,ebi96,gie97}. The soft
state is dominated by a strong thermal component below $\sim3$ keV,
combined with a power-law continuum extending above 1 MeV. The iron
line is visible also in the soft state \citep[e.g.][]{gie99}. The soft
radiation is expected to come from an optically thick, geometrically
thin accretion disc, while the hard emission is usually explained by a
hot, optically thin medium where the soft disk photons are Comptonized
\citep{sha76,pou01}.

We used the hardness ratios (or more specifically count ratios) between
the 9-20 keV and 2-4 keV channels to provide a means of correlating
the PDS properties with changes in the energy spectrum. To compute a
hardness ratio, the 16~s lightcurves were first divided according to
the start and stop times of the data used in each PDS. A mean count
rate was then calculated in each energy range for each interval, and the
hardness ratio $HR$ computed. This way, each PDS could be tied
to an average count rate and hardness ratio. 

\begin{figure}
\resizebox{\hsize}{!}{\includegraphics{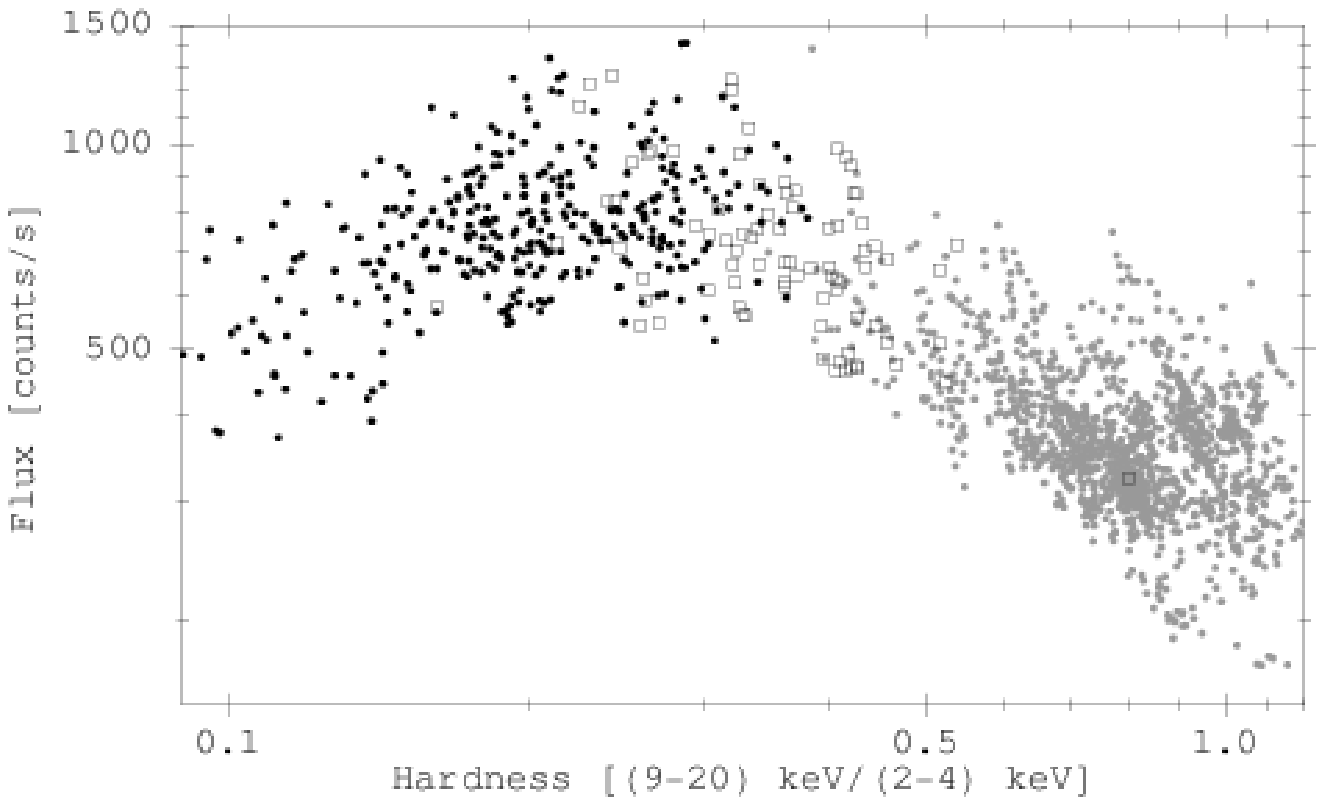}}
\caption{Flux-hardness correlation of Cygnus~X-1. Each point
represents the mean values of hardness and flux during one PDS. Grey
points are the hard state, black points the soft state and open squares
points from transitions and flares. The change in correlation around 
$HR\sim 0.3$ is clearly visible. The flux is the sum of the two channels.}
\label{fluxhard}
\end{figure}

Figure~\ref{fluxhard} shows the flux-hardness relation. In Cyg~X-1, the
correlation between flux and hardness differs between the states. During the 
hard state it is generally negative, while in the soft state there is a
positive correlation \citep{wen01,zdz02}. There is a distinct peak around 
$HR \sim 0.3$, indicating a rather sharp change in behavior.

\begin{figure}
\resizebox{\hsize}{!}{\includegraphics{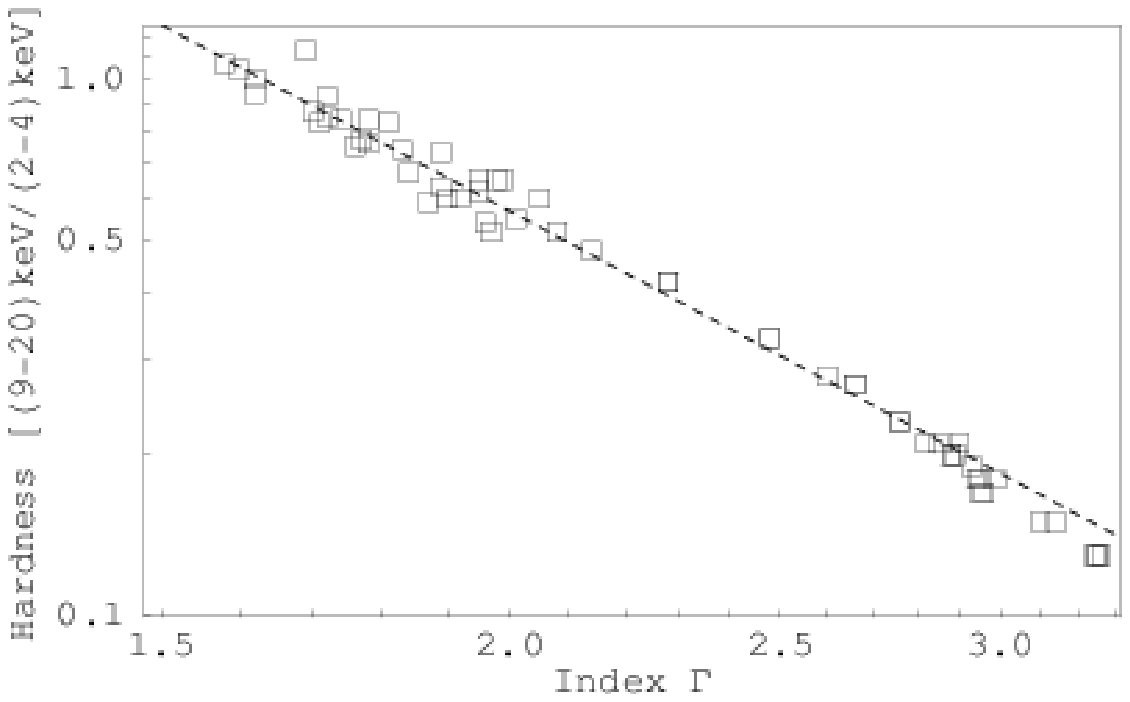}}
\caption{Empirical relation between spectral index $\Gamma$ and 
hardness $HR$, obtained by fitting the energy spectra between 2-20 keV 
with a power-law. Errors are within the boxes. The equation of the
power-law is $HR=3.9\times \Gamma^{-2.8}$.}
\label{indexcorr}
\end{figure}

The hardness as defined above is often a useful tool, as it
provides an easy way to measure spectral changes with very good
time resolution. However, it cannot readily be compared with hardness
values from other instruments or energy bands. 
To facilitate comparison with other studies, we derived an empirical
relation between our definition of hardness ratios and the spectral 
index $\Gamma$. This was done by modeling a number of energy 
spectra in the 2-20 keV range with a power-law in \texttt{xspec}, as 
well as calculating their average hardness. While the energy spectrum of 
Cyg~X-1, especially in the soft state, cannot accurately be 
described by a simple power-law, this approach provides a good approximation
of the spectral characteristics of the source. The resulting empirical 
relation between the index $\Gamma$ and hardness ratio $HR$ was found 
to be well described by a power-law (see Fig.~\ref{indexcorr}).

\begin{equation}
HR=3.9 \times \Gamma^{-2.8}
\label{gammaeq}
\end{equation}

The transitional state of Cyg~X-1 has a typical index in the range 
$2.1 \la \Gamma \la 2.4$ in the energy band 3-12 keV 
\citetext{\citealt{zdz02}; Fig.~3 in \citealt{zg04}}. Comparing to the 
empirical relation of Eq.~\ref{gammaeq}, this would correspond to a 
hardness $0.3\la HR \la 0.5$, with the lower edge matching the 
peak in Fig.~\ref{fluxhard}.
The simplicity of the relation in Eq.~\ref{gammaeq} means that
spectral index and hardness ratios can be used interchangeably, and we
will consistently use the latter in this paper.  

\section{Results}
\label{results}
In this section we present the results of our analysis. We begin with
a general description of how the PDS evolves during spectral transitions,
and then describe in detail the relevant parameter correlations. 

While narrow, well-defined peaks in the power spectrum are generally 
called QPOs, there is no set definition. This leads to a sliding 
scale where wider features are sometimes referred to as breaks and 
sometimes as QPOs. To avoid confusion and connection with the different
types of physical processes associated with breaks and QPOs, we will 
consistently use `temporal feature' or `characteristic frequency' to 
describe the features seen in the PDS, and in the context of our modeling 
components refer to them as 
`Lorentzians'. For a review of temporal features in both BHC and
NS systems, see \citet{vdk95,vdk04}. A possible identification of the 
different features in BHC in terms of those found in NS is presented 
by \citet{mkw04}.  

\subsection{PDS of the different states}
\label{states}
As Cyg~X-1 spends most time in the hard state, the majority of 
the observations were made during this state. It is well described by
two Lorentzian profiles and no power-law component (see Fig.~\ref{model2ex}). 
The two peaks shift in frequency, moving on timescales of hours. In 
very few ($\la 1\%$) of the hard state cases, model 1 did not provide a good 
fit due to a third component being evident at higher frequencies. These 
cases correspond to the observations with highest hardness, and will 
be discussed further below. 

During larger flares and transitions, the two peaks shift to
higher frequencies and it becomes necessary to include the power-law 
component to achieve an acceptable fit. The
power-law component only significally contributes at lower frequencies
(Fig.~\ref{model3ex}). When analyzing the data we separated flares,
transitions from hard to soft state, and transitions from soft to hard
state in an attempt to detect possible differences between them. However, 
in the analysis done in this paper the three appear identical. Our results 
on this point agree with those found previously 
\citetext{e.g. \citealt{pot00,cui02}; \citetalias{pot03}}.

It is during the transitions and flares that the power spectrum of Cyg~X-1 
shows the most rapid changes. We are able to see changes in the PDS 
down to our limit of $\sim 15$ minutes, justifying our choice of
a short integration time in the analysis.

In the soft state the power-law component dominates with an index
close to $-1$. However, in the majority of the cases ($\sim65\%$),
one or two Lorentzian components are needed in addition to the power-law
in order to fit the PDS. The Lorentzian components are particularly 
frequent in the 2001/2002 extended soft state. During this period, the
source made at least two brief transitions down to the hard state 
\citepalias[also noted by][]{pot03}. No trace of these transitions 
is seen in the ASM lightcurve of Fig.~\ref{asmlc}, but both the 
hardness and PDS reverted to that of the canonical hard state.  

\subsection{Relations between the peak frequencies}
\label{relations}

\begin{figure}
\resizebox{\hsize}{!}{\includegraphics{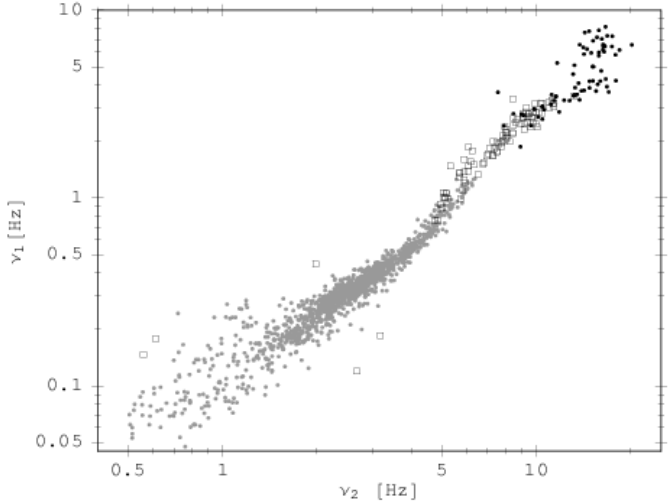}}
\caption{Correlation between the peak frequencies of $L_1$ and $L_2$.
The gray and black points represent the hard and soft states respectively,
and the open squares are the transition state. The break away from the
hard state behavior as Cyg~X-1 enters the transitional state is 
clearly visible.}
\label{parcorr1}
\end{figure}

The relation between $\nu_1$ and $\nu_2$ is shown in Fig.~\ref{parcorr1}.
Our observations confirm the correlation between the peak frequencies
of the PDS found in previous studies of Cyg~X-1 in the hard state 
\citepalias{now00,pot03}. Since our study is based on a much larger
data set, we can make a more accurate analysis of the hard state 
frequency variations. We are also  
able to add the behavior of the peak frequencies during transitions, 
flares and into the soft states. A clear feature in Fig.~\ref{parcorr1} is 
the steepening of the relation at $\nu_2 \simeq$ 5 Hz. This corresponds to 
the region where the source goes from hard state into the
intermediate/transitional state. We note that this is the region
where \citetalias{pot03} see a shift in power from their lower frequency
Lorentzian ($L_1$), to a higher frequency one ($L_3$). Since we
use only two Lorentzians, the break could in principle be attributed
to a change in the components being fitted. We have however carefully
monitored this region, and find no evidence of such a change in our
data. Since our method of analysis allows us to detect changes in the PDS 
on timescales as short as $\sim 15$ minutes, it is more likely that the 
break is due to a real change in the behavior of the components. The 
fact that the points lie in a continuous distribution, both in 
Fig.~\ref{parcorr1} and in the following figures, further supports 
this identification.

In the soft state, the relation between the two peak frequencies is
somewhat uncertain. We see two major effects behind the large spread
in these data. Since we do not correct the Poisson noise level for 
dead-time effects, our frequency range is limited to the region where
these can be neglected. As the second 
Lorentzian approaches our upper frequency limit of 25 Hz, it 
becomes difficult to constrain. Even with this rather conservative limit,
we cannot rule out that the points near the limit are affected by dead-time
effects. A second effect which further complicates the modeling is the 
fact that the power of the feature is reduced at higher frequencies. As we 
average over only ten power spectra, our signal-to-noise ratio does not 
allow us to model weak components. The spread of points in the soft state is 
therefore due to both the limitations set by dead-time effects and the 
signal-to-noise ratio. 
 
\subsection{Comparing the two Lorentzians}

\begin{figure}
\resizebox{\hsize}{!}{\includegraphics{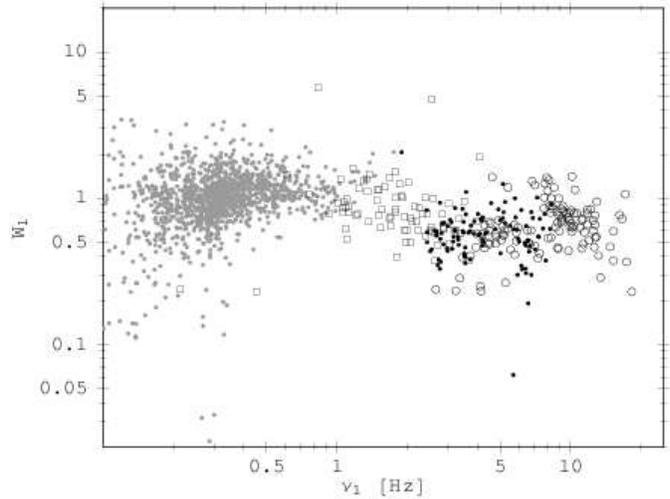}}
\caption{Relation between $W_1$ and $\nu_1$. Symbols are the same as
in Fig.~\ref{parcorr1}, with open circles being points from the single
Lorentzian in model 3. The width is generally constant throughout the
frequency range.}
\label{parcorr3}
\end{figure}

\begin{figure}
\resizebox{\hsize}{!}{\includegraphics{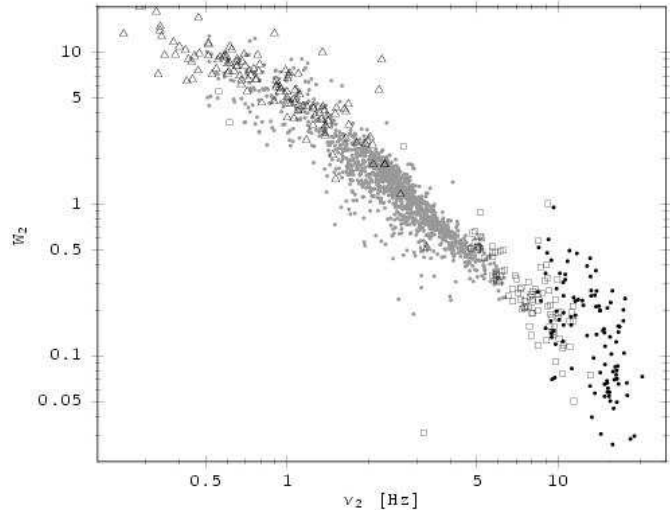}}
\caption{Relation between $W_2$ and $\nu_2$. Symbols are the same as
in Fig.~\ref{parcorr1}, with open triangles being points from the 
hard state where the relation between $\nu_1$ and $\nu_2$ was fixed. The
negative correlation between $W_2$ and $\nu_2$ is clear, and can 
approximately be described by a power-law with index $\sim-1.5$.}
\label{parcorr2}
\end{figure}

The strong correlation between the peak frequencies of the two Lorentzians 
in Fig.~\ref{parcorr1} suggests a close physical link between the components. 
However, when studying the relations between $W_i$ and $\nu_i$ the two 
components show completely different behavior (Figs.~\ref{parcorr3} and 
\ref{parcorr2} respectively). The first Lorentzian shows no 
correlation between these parameters, and $W_1$ remains fairly constant.
For the second Lorentzian, $W_2$ changes roughly two orders of magnitude,
and there is a strong correlation, which can approximately be
described by a power-law with index $\sim-1.5$. In contrast with 
the relation in Fig.~\ref{parcorr1}, the points for the transitional 
and soft states seem to follow the same relation as in the hard 
state. 

During the hard state there are cases where the Lorentzians
shift to very low frequencies. These cases correspond to observations 
with very low flux and a high hardness ratio. To provide a good fit in 
these cases, we found it necessary to freeze the ratio between the peak 
frequencies by extrapolating the relation found at higher frequencies 
(see Sect.~\ref{modelcomp} below). For these low frequencies in the hard 
state, there is a hint of flattening of the index in the $W_2$-$\nu_2$
correlation (open triangles). Since this occurs in the region where $\nu_1$ 
was fixed in relation to $\nu_2$ we cannot rule out that this is an 
artifact of the fitting procedure. 

In Fig.~\ref{parcorr3}, the open circles represent the single Lorentzian 
of model 3 (see Table~\ref{components}). We identify this component as
$L_1$. The identification is supported by the fact that the points from
this component well matches the other soft state points, both in 
this and subsequent figures. If the component is instead identified as
$L_2$, the distributions occupy completely separate regions in the 
parameter space. We surmise that as both Lorentzians move up in frequency,
eventually $L_2$ becomes too weak and moves too high in frequency for
us to see in our analysis, and only $L_1$ is visible.      

\begin{figure}
\resizebox{\hsize}{!}{\includegraphics{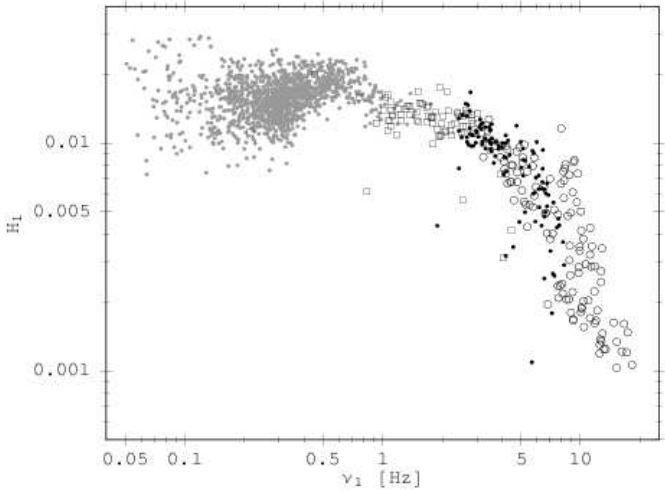}}
\caption{Power at the peak frequency versus peak frequency
for $L_1$. Symbols are the same as in Fig.~\ref{parcorr3}. The
change in behavior during and after the transition (open squares)
is rather gradual.}
\label{parcorr6}
\end{figure}

\begin{figure}
\resizebox{\hsize}{!}{\includegraphics{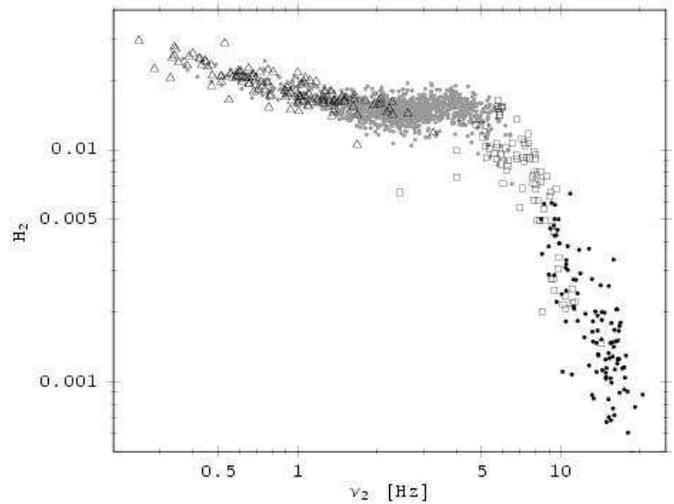}}
\caption{Power at the peak frequency versus peak frequency
for $L_2$. Symbols are the same as in Fig.~\ref{parcorr2}. Note the
sudden change in behavior as the source enters the transitions (open 
squares). This suggests the underlying physical mechanism reacts 
sharply to changes, e.g. in the accretion geometry, occurring at 
these periods.}
\label{parcorr7}
\end{figure}

Comparing the relations between $H_i$ and $\nu_i$ shows similar 
behavior for $L_1$ and $L_2$ (Figs.~\ref{parcorr6} and \ref{parcorr7}).
Both components stay at roughly constant height at lower frequencies,
and decrease rapidly with frequency at higher $\nu_i$. This change
in behavior for $L_2$ occurs as Cyg~X-1 enters the transitional 
state, suggesting that the mechanism behind $L_2$ starts to weaken as the
transition begins, perhaps due to a change in the accretion geometry.
We surmise that the behavior of $H_1$ at lower frequencies
is most likely explained by the effect first noted by \citet{bh90} in 
their analysis of the PDS in terms of a doubly broken power-law model. 
They found a negative correlation between the frequency of the first 
break and the power at that frequency. Transforming this correlation into
the $fP_f$ representation leads to approximately constant $H_1$ as a 
function of $\nu_1$. Our results are also consistent with the relation 
between normalization and peak frequencies of the Lorentzian components
seen in Fig.~6 of \citetalias{pot03}, but note the difference in parameters 
used and identification of components. 

Combining the behavior of $H_i$ and $W_i$ once again shows the difference
between the two components. Since both $H_1$ and $W_1$ stay roughly
constant at low $\nu_1$, the contribution from $L_1$ to the total
RMS variability remains the same, weakening only above $\sim 5$ Hz, when
the source is in the soft state. In contrast, the RMS contribution from
$L_2$ falls with $\nu_2$ over the whole frequency range. As $H_2$ 
starts to decrease, $L_2$ weakens very rapidly. Interesting is also
that the rapid decrease in height starts at $H_i\sim 5$ Hz for
both $L_1$ and $L_2$. A possible explanation for the similarities and
differences in behavior of the components will be presented in 
Sect.~\ref{behavior}.

\subsection{Evolution of the PDS}
\label{pdsevol}

\begin{figure}
\resizebox{\hsize}{!}{\includegraphics{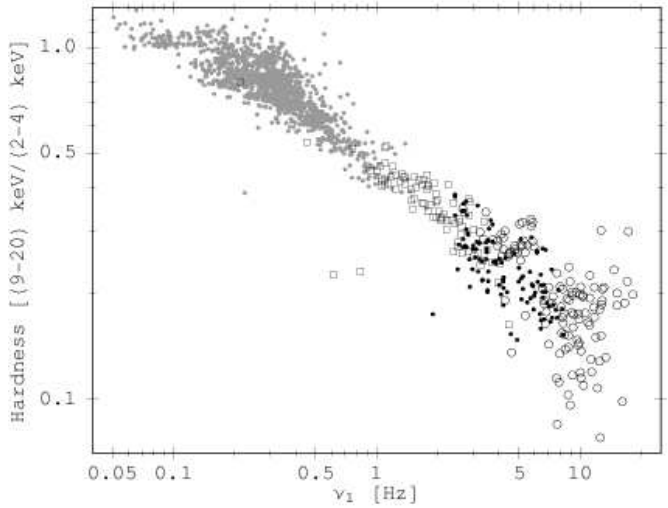}}
\caption{Relation between hardness and $\nu_1$. The symbols are the
same as in Fig.~\ref{parcorr3}. There is no apparent change in behavior
as the source enters the transitional or soft states.}
\label{parcorr4}
\end{figure}

\begin{figure}
\resizebox{\hsize}{!}{\includegraphics{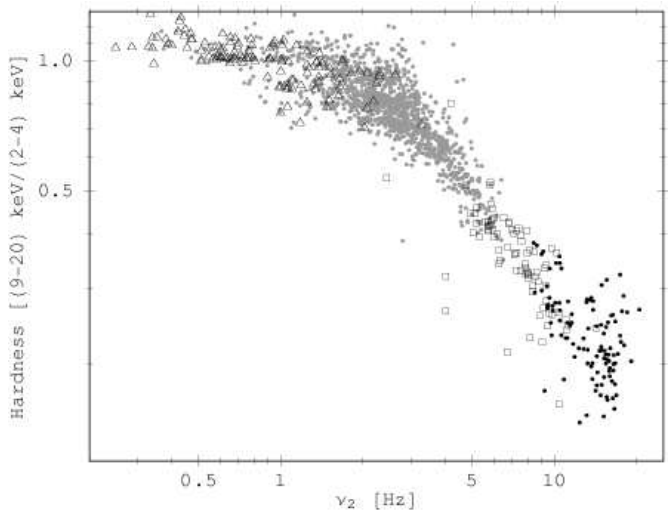}}
\caption{Relation between hardness and $\nu_2$. The symbols are the
same as in Fig.~\ref{parcorr2}. Note the tendency for the hardness to
`flatten out' at $\sim 1.1$.}
\label{parcorr5}
\end{figure}

As described above in Sect.~\ref{states}, there are systematic
changes in the PDS tied to the changes in hardness. Figures~\ref{parcorr4} 
and \ref{parcorr5} show the evolution of $\nu_i$
with hardness. Both peak frequencies increase with decreasing hardness.
This movement follows approximately the same pattern
in all states while the components are detectable, which we take 
as another indication that we are following the same two components in
all cases. We note that this gradual increase in peak frequency 
would be expected if the Lorentzian components are assumed
to arise in a region that is closer to the central object in the
soft state than in the hard state (e.g. the inner edge of a truncated
accretion disk or a transition region between disk and corona). 
Such models have been presented by e.g. \citet{pou97}, \citet{chu01},
and \citet{zdz02}. 

Beyond the necessity to include the power-law
component in our models below a hardness of $\sim 0.4$, 
Figs.~\ref{parcorr4} and \ref{parcorr5} show no change in behavior
around the transition phase hardness, $HR\sim 0.3$ seen in 
Fig.~\ref{fluxhard}. We therefore conclude that the changes responsible 
for altering the flux-hardness correlation do not affect the movement of 
the peak frequencies. 
The other parameters do not show any evolution with hardness, 
beyond a trend for the power-law to extend to higher frequencies
as the hardness decreases. This is a natural effect of the
model as the Lorentzian components move away from the lower frequencies. 
The behavior is however consistent with the 
$f^{-1}$ power-law gradually extending higher in frequency as the 
source goes from hard state to soft state \citep{zdzprep}.

When determining the state of the source, we have done so from the 
point of view of the PDS, as they are the base of our analysis.
Figures \ref{parcorr4} and \ref{parcorr5} show that there is a unique
relation between a point in the PDS parameter space and the hardness.
We are therefore able to compare the `PDS states' with definitions 
based on the spectral index of the source. We see that our transitional 
state PDS roughly fall in the range $0.3\la HR \la 0.5$. Below this range 
is the soft state and above the hard state. From the relation in 
Eq.~\ref{gammaeq}, the corresponding range in spectral index is 
$2.1\la \Gamma \la 2.4$, with the hard state having a lower and the soft 
state a higher index. We note however, that these boundaries are not 
strict, and that there is some overlap between the different states.
This range matches well with previous results, e.g. \citet{zdz02,zg04}.

\begin{figure}
\includegraphics[width=88mm,height=0.9\textheight]{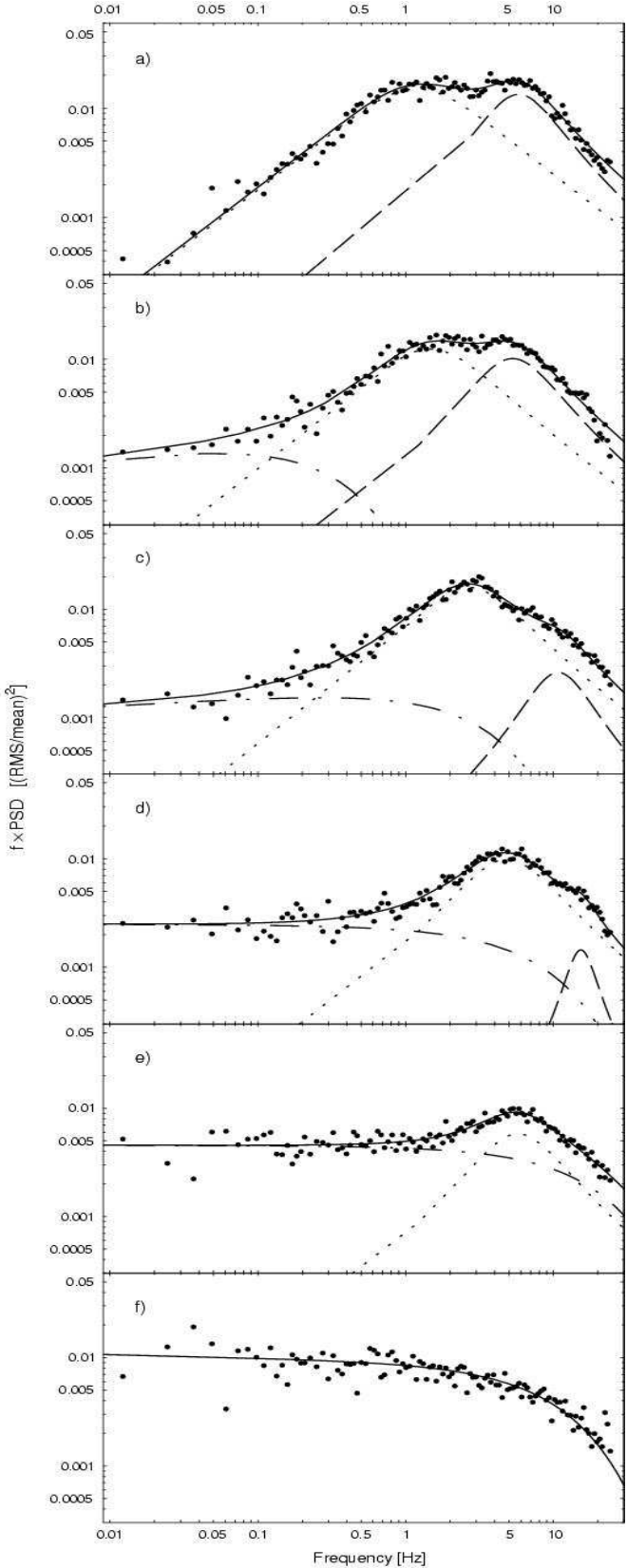}
\caption{Montage showing the evolution of the PDS of Cyg X-1 from
hard state (panel \textbf{a}) to canonical soft state (panel \textbf{f}). 
The top three panels (\textbf{a-c}) are from observations made within a 
few hours during MJD 52\,330, and clearly illustrate the changes that 
can occur on this timescale. Panels \textbf{d} and \textbf{e} are from 
MJD 52\,324, and are also no more than a few hours apart. The bottom panel 
(\textbf{f}) is from an observation
on MJD 52\,358. Note the weakening and narrowing of the upper frequency
Lorentzian (dashed line) as it shifts to higher frequencies.}
\label{evolution}
\end{figure}

To illustrate the complete evolution of the PDS, Fig.~\ref{evolution}
presents a series of fittings showing the gradual change from 
normal hard state to canonical soft state PDS. Unfortunately, the
observations made did not capture the complete evolution in a chronological
series, so Fig.~\ref{evolution} is a montage of PDS from observations made
during a brief transition to the hard state in the extended soft state of 
2001/2002. The top three panels are from observations made on MJD 52\,330 
and panels \textbf{d} and \textbf{e} are from an observation made on 
MJD 52\,324. The last PDS (panel \textbf{f}) is made on MJD 52\,358, when 
the source had once again returned to the soft state. Figure~\ref{evolution} 
illustrates the shift in the peak frequencies of the Lorentzians, and the 
gradual increase of the power-law as the source goes from hard state to 
soft state.  

\section{Discussion}
\label{discussion}

The results of our analysis show that the PDS of Cyg~X-1 
can be successfully modeled in all the spectral states using only a 
few components. The observed behavior of the Lorentzian components
lends weight to the argument that they indeed represent the temporal
characteristics of the physical processes giving rise to the PDS.   

\subsection{Identification of components}
\label{identification}

As noted in Sect.~\ref{relations}, the identification of components in 
the PDS is not trivial. \citetalias{pot03} in their study use four Lorentzians
to fit the power spectrum. While their fourth component, $L_4$, is
above our upper frequency limit, our model will fit the two dominating
components at a given time. According to the identification of 
\citetalias{pot03}
this means that in the canonical hard state we fit their first two 
Lorentzians, but in the larger flares and transitions we fit their 
second and third Lorentzians. The first Lorentzian is then quite weak
and will be `absorbed' in our power-law component. Supporting this
identification is the change in the correlation of the peak frequencies
during transitions and large flares. 

Recently, \citet{vst04} has argued that the identification of 
\citetalias{pot03} leads to significant
deviations during `failed state transitions'. For canonical hard state
PDS, the first two Lorentzians dominate and follow the relation of 
Fig.~\ref{modelpic}, but \citet{vst04} points out that if the second
and third components dominate in transitions and large flares, the 
peak frequencies of the first two components deviate significantly
from the correlation in the normal hard state. If instead the dominating 
components are still identified as the first two Lorentzians, the peak 
frequencies will follow the trend seen in Fig.~\ref{modelpic}. Although the 
points do not lie along the extension of the power-law described at lower 
frequencies, their behavior matches that of the other sources, as seen 
in \citet{wvdk99}. To support the identification, \citet{vst04} adds data 
from a flaring episode in November 2000, but a gap between the 
canonical hard state PDS and these flares still remains.  

Since the study reported here allows for shorter timescales to be 
monitored, we are able to follow changes in the Lorentzians in great
detail. From Fig.~\ref{evolution} it
is clear that our model can explain the evolution of the PDS, and provide
extensive coverage of all states of the source. As 
seen in Fig.~\ref{parcorr2} our results show a continuous distribution 
between the hard state and transitions, and we are able to fill in the 
gap in the data of \citet{vst04}. The results here therefore favor this 
identification over that of \citetalias{pot03}.    

\subsection{Comparisons with models}
\label{modelcomp}

The temporal features observed here have been detected
both in NS and BHC systems. The origin of these 
characteristic frequencies has been long debated in the literature, and 
various models have been proposed in each physical scenario \citep[for 
an extensive review, see][]{vdk04}. However, 
recent comparative studies between these two types of sources have shown 
that they present very similar frequency correlations, over a
frequency range of many decades \citep{pbk99,bel02}. The similarities
suggest the same underlying physical mechanism, which cannot be dependent
on either a magnetosphere, a solid surface, or an event horizon 
\citep{wij01}. 
Beat-frequency models, which have been extensively used in the context 
of NS, can not easily be extended to black holes as these are unable to
provide an anchor to the magnetic field, a condition that is needed to 
generate a beating with the Keplerian frequency. 

Several classes of models have been proposed to explain the
observed QPO's and their correlations \citep[see][~and references therein]{pbk99}. They commonly ascribe the variations to processes in the accretion
disk. This can be done by identifying observed frequencies with orbital, 
epicyclic and precessional frequencies at certain radii (such as the
inner edge of the disk). Another approach is to study disk oscillation
modes. For example, in the region close to the compact object, global 
disk oscillation modes can become trapped in particular disk annuli.
For comprehensive summaries on this topic, see e.g. \citet{wag99} and
\citet{kat01}. 
As demonstrated in \citet{wag01}, disk mode frequencies are mainly 
determined by the mass and angular momentum of the compact object, and 
depend only weakly on the disk parameters. While this property makes the
models good candidates in explaining the high coherence seen in higher 
frequency QPO's \citep{bar04}, it also makes it difficult to explain
the large (over one order of magnitude) shifts in peak frequency observed
in our study. In addition, the features we model with Lorentzians have low
coherence. Our parametrization does not directly give a value for the
more generally quoted quality factor $Q=\nu/\Delta \nu$, but for $L_1$ 
this value is $\la 1$ in all states. For $L_2$, $Q$ varies with the
spectral state. It is low ($Q<1$) in the hard state and while higher 
in the soft state, $Q<5$ in all observations.  

\begin{figure*}
\centering
\includegraphics[width=17cm]{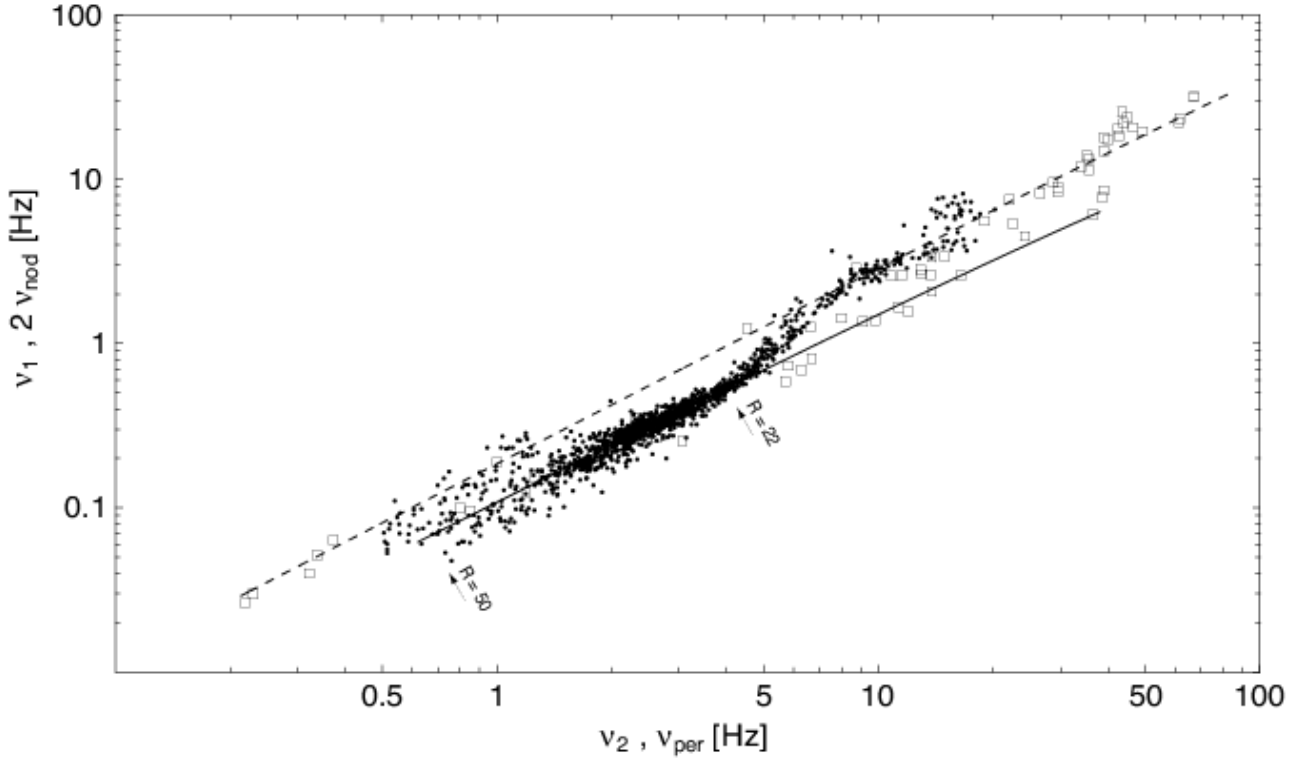}
\caption{Relation between peak frequencies (black points) together 
with the fit to the relation predicted in the relativistic precession 
model of \citet{SV98,SV99}. Also plotted are the points from Fig.~2a in 
\citet{wvdk99} (open boxes), with data from both neutron star (atoll 
sources) and BHC sources. The arrows mark different inner radii of 
the accretion disk, in units of gravitational radii. The solid line is 
the fit to the 1230 hard state points in the case of prograde rotation, 
when $\nu_1 = 2\nu_{\rm nod}$. The best-fit power-law has an index of 
$1.20\pm 0.01$. As the source enters the transitional state, the
relation clearly breaks away from that predicted by the RPM model. The
dashed line corresponds to a power-law of index 1.20 shifted to approximately
follow the points from the soft state.}
\label{modelpic}
\end{figure*}

In the following, we will consider the somewhat idealized case of the 
relativistic precession model \citep[RPM,][]{SV98,SV99}. It states that 
the timing features are the observable effect of the relativistic nodal 
and periastron precessions of a disk element or a `blob' in a slightly
tilted and eccentric orbit, respectively. Equally possible is the 
interpretation of the features arising in a narrow band or annulus in
the inner region of the accretion disk \citep[see e.g.][]{vdk04}. Although 
the predicted frequencies
are those for free particle orbits, these can still often be recognized 
in the models of disk modes \citep[e.g.][]{wag99}. \citet{psa00} showed that 
a sharp transition region in the accretion disk acts as a low-band pass 
filter with strong resonances near the frequencies predicted by the RPM, 
motivating a closer look at this model. \citet[][ hereafter
SVM]{SVM99} have shown good agreement between the predicted
frequency correlations in the RPM model and a small number of BHC and
NS observations. Although the model currently requires NS spin rates and 
masses higher than measured, these problems are dependent on the chosen 
equation of state, and the RPM model successfully predicts
the observed quadratic dependencies between lower and higher frequency 
features seen in many sources \citep[see][ and references therein]{vdk04}. 
Based on previous calculations
\citep[by][]{BPT72,OKF87,Kato90} of the Keplerian $\nu_\phi$,
epicyclic $\nu_r$, and vertical $\nu_\theta$ frequencies for a Kerr
black hole, \citetalias{SVM99} derived that in the weak field approximation 
for a slowly rotating black hole
\begin{equation}
\nu_{\rm nod} \propto a_*  M^{1/5} \nu_{\rm per}^{6/5} \; ,
\label{rpm}
\end{equation}
\noindent where $\nu_{\rm nod}=\nu_\phi-\nu_\theta$ is the nodal
precession frequency, $\nu_{\rm per}=\nu_\phi-\nu_r$ is the periastron
precession frequency, $M$ is the mass, and $a_*=a/R_{\rm g}$ is the
dimensionless specific angular momentum (with $a=J/M c$, $R_{\rm g}=G
M/c^2$ being the gravitational radius, $J$ the angular momentum, $c$
the speed of light in vacuum, and $G$ the universal gravitational
constant). We used a power-law model to fit the correlation between
frequencies for the hard state data (see Fig.~\ref{modelpic}), up to the
{\it break} observed in the transitional state at $\nu_2 \simeq
5$ Hz. The best-fit parameters did not change significantly selecting a
somewhat smaller, more conservative upper limit, nor was the fit affected
by ignoring the points at the lowest frequencies. We obtained a power-law 
index $1.20\pm0.01$ in remarkably close agreement with Eq.~\ref{rpm}.

In \citetalias{SVM99} it is argued that due to the geometry of a
tilted inner disk, the second harmonic of the nodal precession
frequency gives the strongest modulation, and that this assumption 
provides the best match for their observations. We also found that a 
good fit of our
data using the $\nu_1=\nu_{\rm nod}$ identification would only be
possible for an unrealistically large black hole mass. Adopting the
\citetalias{SVM99} identification, i.e. in our case $\nu_1=2\nu_{\rm
nod}$ and $\nu_2=\nu_{\rm per}$, and using the full relativistic
expressions for the precessional frequencies, it is possible to
derive some of the physical parameters of the system. For this
purpose we used the $M = 8 M_{\sun}$ mass estimate \citep[see][~and 
references therein]{now99}. Although this $M$ value is somewhat uncertain, 
the frequency relation depends weakly on the black hole mass as shown by
Eq.~\ref{rpm}. At this frequency range, there are two possible
solutions for the specific angular momentum $a_*$, depending on whether
the orbital motion is prograde or retrograde. We obtained
$a_*=+0.49\pm0.01$ and $a_*=-0.57\pm0.01$ respectively, comparable to
previous results based on modeling of energy spectra 
\citep[see e.g.][]{ShL76,ZCC97}. In
Fig.~\ref{modelpic} the best fit for the prograde case is shown ({\it
solid line}), and some derived orbital radii are indicated.
During the hard state, within the assumptions of the RPM, the
inner radius $R_{\rm in}$ of the disk ranges from 50 to
22 $R_{\rm g}$ when reaching the transitional zone. If the retrograde
case is instead assumed the corresponding values change very little,
with the inner disk ranging from 55 to 25 $R_{\rm g}$. These ranges are 
in very good agreement with coronal model estimates for Cyg~X-1
\citep[see][~and references therein]{MP99}.

The behavior during the transitional and soft states seems to be more
complex, in part because of the spread of the data points at the
highest frequencies. If all the data above the break is modeled with
a second power-law (thus given an index $1.7\pm0.03$) the fit
residuals clearly do not appear randomly distributed. Another
possible interpretation of the data is that after the transition, the 
points corresponding to the soft state follow a power-law with an index
identical to that of the hard state. This picture is reinforced when 
comparing our Cyg~X-1 RXTE data to those presented in 
a study of the frequency correlations in X-ray binaries by \citet{wvdk99},
from many different NS and BHC sources (open squares in Fig.~\ref{modelpic}). 
The data points seem to lie along `parallel tracks' as predicted by the
RPM. This behavior, along with the gradual shift between the two 
tracks during the transitional state, must then be explained. It seems 
likely that a reconfiguration of the accretion disk geometry during the 
state transitions is responsible for the changes in the relation between the 
frequencies.  

A suggestion to explain the state transitions in Cyg~X-1,
first made by \citet{ShL76} and later on by \citet{ZCC97}, is
that they are produced by the reversal of the disk rotation. This
would occur due to the unstable nature of the stellar wind from the
primary star feeding the accretion. In this scenario, the system would
switch back and forth from a retrograde rotation in the hard/low
state, to a prograde rotation in the soft/high state. Differences in
luminosity would be explained by the changes in the `last stable
orbit' $R_{\rm last}$, that in the case of maximal rotation
would change from $R_{\rm last}(a_*=-1)=9 R_{\rm g}$ to $R_{\rm
last}(a_*=+1)=R_{\rm g}$.  According to our previous fit of the hard
state data, a reversal of the disk rotation would imply that
$a_*=+0.57$ in the soft state. However, this value does not fit the
soft state data. \citetalias{SVM99} showed that corrections for 
non-negligible eccentricities or tilt angles are generally small. Very
large eccentricities ($e > 0.85$) would need to be invoked to explain
the data. The rotation reversal scenario could thus only be maintained
with additional assumptions.
   
Reducing the spread of points in the soft state data could help
determine the behavior of the relation and thereby further constrain
models. A more thorough study of the soft state data, extending to higher 
frequencies, is currently underway. This analysis will also make use of 
data recently made public from an additional soft state in June/July 2003. 

\subsection{Behavior of components}
\label{behavior}

As noted by \citet{bel02}, when using Lorentzian components 
to model both broad and narrow features in the power spectrum, the
peak frequency $\nu_i$ does not have an immediate interpretation in
the picture of a damped harmonic oscillator. Interpreting the features
as arising by superpositions of a range of frequencies (such as from
an annulus in the innermost region of the accretion disk), $\nu_i$
instead becomes a measure for the highest frequency contributing, or
the one contributing the most power. Our width parameter $W_i$ can
then be seen more as a measure of the extent of the contributing 
region, than as a measure of the lifetime of an individual 
process. For a deeper discussion on this topic, see \citet{bel02}.  

The different behaviors of the width parameter $W_i$ in the two
components directly rule out the simple case of the components
originating in the same `blob'. If the Lorentzians instead arise in an
annulus close to the inner radius $R_{\rm in}$ of the accretion disk,
increasing $\nu_i$ indicate decreasing $R_{\rm in}$. The decrease of
$W_2$ with frequency may then be understood by a more narrow annulus
dominating the radiation at smaller $R_{\rm in}$. However,
$W_1$ is roughly constant as $\nu_1$ increases. If the two components
arise in the same annulus, $W_1$ may be expected to behave in the same
way as $W_2$. As can be seen in Figs.~\ref{parcorr3} and
\ref{parcorr2}, this is not the case. Any model in which the two
components arise from the same annulus then requires a mechanism
allowing $L_1$ to show different behavior than $L_2$. In the context
of the RPM, where $L_1$ is identified as the nodal frequency, a
possible explanation is that warping of the disk leads to a smaller
deviation from the orbital plane at smaller $R_{\rm in}$. This would
lead to the modulation decreasing with decreasing $R_{\rm in}$, thus
compensating for the dominance of the innermost ring in terms of
emitted radiation. The result would be a larger `effective' annulus
contributing to $L_1$ than to $L_2$, which is dominated by the narrow
innermost region. For $L_1$, the net effect must be an approximately
constant width as seen in Fig.~\ref{parcorr2}. We note that during the
hard state, $R_{\rm in}$ decreases by about a factor of 2
(Fig.~\ref{modelpic}), while the flux increases with approximately the
same factor (Fig.~\ref{fluxhard}), suggesting a $\sim 1/R_{\rm in}$
relation. If the deformation of the disk scales as $R$, the net effect
would be consistent with a constant power of the component as seen for
$L_1$. If this simple picture is true, $L_1$ could provide important
constraints on the geometry of the accretion disk close to $R_{\rm
in}$.
 
The argument used for the different behaviors of $W_1$ and $W_2$ can 
also be extended to the case of $H_i$ during transitions 
(Figs.~\ref{parcorr6} and \ref{parcorr7}). $H_2$ starts decreasing
rapidly as the transition period starts, whereas $H_1$ shows a more
gradual reaction. If the transitional stage 
is considered to be a phase of rapid changes in the accretion geometry, 
a feature arising from a narrow `ring' will react more sharply than 
one originating in a wider disk annulus. After the transition both 
Lorentzians show the same behavior, indicating that the geometry once
again becomes more stable. 
Since the normalization used gives the relative power of different 
components, it is natural that as the Lorentzians shift to higher 
frequencies, the power-law component will start to dominate the PDS. 
The decrease in power of the Lorentzians will then mainly be a decrease 
in relative strength, rather than an intrinsic weakening of the physical 
processes giving rise to the components. 
We note that this scenario is also a possible explanation for the 
changes in the frequency correlation in Fig.~\ref{modelpic} during
transitions. During the transition the changes in accretion geometry
disturb the relation between the peak frequencies, but as the source
settles into the soft state the correlation is resumed. An explanation
must still be found for the difference between hard and soft state, e.g. 
large deviations from a circular orbit or a shift from the second to  
higher harmonics.

The decrease of $W_2$ with peak frequency is also interesting when
comparing narrow QPOs with broader features (such as 
breaks or `humps'). \citet{pbk99} observe that whereas so-called lower 
kHz QPOs seen in neutron star sources are narrow well-defined peaks, 
components identified as the lower kHz QPO at much lower frequencies 
are broader. As our observations of $L_2$ indicate, studies such as 
this one have the potential to show the complete evolution from the 
broader features at low frequencies to narrow QPOs at high frequencies. 
Tracking these components will determine whether they indeed originate 
from the same process.

A feature seen especially in Fig.~\ref{parcorr5}
is the tendency for the hardness to `flatten out' at a value 
$HR\sim 1.1$. Cyg~X-1 only reaches these levels of hardness in
observations with very low flux. The behavior of the hardness
at these times could be explained by the index of the power-law 
component in the energy spectrum ($\Gamma$) reaching a minimum value, 
and changes in flux (and thereby the PDS) arising from another component, 
e.g. small flares. If the variations of $\Gamma$ are attributed
to variations in the accretion disk rather than the hot inner flow/corona
\citep[e.g.][]{zdz02}, the `flattening' of the hardness could be seen 
when these variation reached some outer boundary, and changes in flux 
were instead due to smaller variations in the processing of the soft 
input photons. That the peak frequencies of the Lorentzian components
also vary at these times might suggest they are related to the
corona rather than the accretion disk. Arguments
for relating `breaks' or QPO features in Cyg~X-1 to a transition 
region at the inner edge of the disk or to the corona have been presented 
by e.g. \citet{chu01} and \citetalias{pot03}. However, we note that this 
behavior can also be explained by variations in the local disk accretion 
rate. Such a mechanism has been suggested by \citet{zdz02} to explain 
variations in amplitude of the hard state energy spectrum with no change 
in shape, i.e. roughly constant~$\Gamma$. 
 
As stated in Sect.~\ref{models}, during the observations with lowest
flux we froze the relation between $\nu_1$ and $\nu_2$ according
to the solid line in Fig.~\ref{modelpic}. The flux of these observations 
were typically below
$\sim 500$ counts/s in the 2-9 keV range, and the hardness was $HR\ga 1$.
The vast majority of these observations were made between MJD 50\,400
and MJD 51\,000 (November 1996 -- July 1998, cf. Fig.~\ref{asmlc}), and 
constitute $\la 10\%$ of the hard state observations.  
During these low flux observations, we could also see an
excess above 10 Hz in almost half the cases. This may well
be a third component, which we only see the times it has moved
to the lower edge of its frequency range. In observations with
a more typical flux level it moves too high in frequency to be detected 
in our analysis. We surmise this is the highest frequency component 
seen by \citetalias{now00}, and $L_4$ in \citetalias{pot03}. This feature 
could add to the spread of points at low $\nu_i$, but does not otherwise 
affect our results.

\section{Conclusions}
\label{conclusion}

After more than eight years of observations, the RXTE data archive
is an excellent source for long-term studies. Using data from this archive
we were able to follow the evolution of the PDS of Cyg~X-1 from 
February 1996 to May 2003. By choosing a simple yet flexible approach for 
our analysis we could model the major features in all states with only
three components. Our method also allowed us to detect changes in
the power spectrum on timescales as short as $\sim15$ minutes.  

Our analysis provides several new results. The main result is 
the successful modeling of the PDS of Cyg~X-1 through all the states.  
We prove that our choice of models and parameters successfully enables 
us to show a natural transition from hard state to intermediate
to soft state and back, and also allows us to study the behavior of the
Lorentzian components in detail. The combination highlights new
correlations between the parameters with interesting physical 
interpretations. 

We find that the spectral state of the
source can be uniquely determined from the parameters of the
Lorentzian components. The parameters used here are the width $W_i$ 
and peak frequency $\nu_i$ of the Lorentzians, along with the power
at the peak frequency, $H_i$. The correlations between parameters can 
all be described by continuous functions, which differ between components. 
This is most evident when comparing Figs.~\ref{parcorr2} and \ref{parcorr3}. 
While the correlations between the peak frequencies and between $H_i$ 
and $\nu_i$ change with the spectral state (cf. Figs.~\ref{parcorr1}, 
\ref{parcorr4} and \ref{parcorr5}), the other correlations show 
no such behavior.

A second result of our study was to show that the PDS of Cyg~X-1
is dominated by the same two Lorentzian components at all times. 
We base this conclusion on the fact that we saw no change in the 
identification of components on our analysis timescale and that the 
points in Figs.~\ref{parcorr1} to \ref{parcorr5} each describe a continuous 
distribution. The relation of the peak frequencies of these components 
follows the pattern seen in both other BHC and NS systems, and we 
therefore conclude that the physical processes responsible for them cannot 
be explained by invoking magnetic fields, a solid surface, or an event 
horizon. A possible model is the RPM of \citet{SV98,SV99}, which describes 
the hard state data remarkably well, but requires some rather {\it ad hoc}
assumptions to fit the soft state. One possibility is to invoke a change
of rotational direction during the state transitions. We find that this 
scenario alone is insufficient to explain the data, and additional
assumptions are still required. Another point that speaks against the 
rotational reversal model for state transitions is the observed similarity 
with other sources, some of which are believed to accrete through Roche lobe 
overflow and thereby not experience any changes in the rotation of the disk.

We also compared the results of transitions from hard to soft, soft
to hard and flares for systematic differences. Within our analysis
we were unable to differentiate between the three, and we cannot
show any way to predict whether such an event will lead to a full 
source transition or a large flare.  

With this study we have shown the importance of looking at the PDS
of Cyg~X-1 in a broader context covering all the states of the source.
To this end we are currently studying the soft state PDS in more detail
and adding newly available data. Determining the behavior of the Lorentzian 
components at higher frequencies will be of great importance for
increasing our understanding of the physical processes behind them. 

\begin{acknowledgements}
We wish to thank Juri Poutanen and Andrzej Zdziarski for helpful 
discussions, and Michiel van der Klis for providing us with the additional 
data in Fig.~\ref{modelpic}. We are grateful to an anonymous referee
whose comments and suggestions helped improve this paper.

This research has made use of data obtained through the High Energy 
Astrophysics Science Archive Research Center (HEASARC) Online Service, 
provided by NASA/Goddard Space Flight Center. 
For the ASM lightcurve we used results provided by the ASM/RXTE teams at 
MIT and at the RXTE SOF and GOF at NASA/GSFC. This work has been partly
funded by the Swedish National Space Board.

\end{acknowledgements}


\begin{thebibliography}{}

\bibitem[Ba\l uci\'nska-Church et al.(1995)]{bal95} 
Ba\l uci\'nska-Church, M., Belloni, T., Church, M.~J., \& Hasinger, 
G.\ 1995, \aap, 302, L5

\bibitem[Bardeen et al.(1972)]{BPT72} Bardeen, 
J.~M., Press, W.~H., \& Teukolsky, S.~A.\ 1972, \apj, 178, 347 

\bibitem[Barret et al.(2004)]{bar04} Barret, D., Kluzniak, 
W., Olive, J.~F., Paltani, S., \& Skinner, G.~K.\ 2004, ArXiv Astrophysics 
e-prints, astro-ph/0412420

\bibitem[Belloni \& Hasinger(1990)]{bh90} Belloni, T.~\& 
Hasinger, G.\ 1990, \aap, 227, L33 

\bibitem[Belloni et al.(1996)]{bel96} Belloni, T., Mendez, 
M., van der Klis, M., et al. 1996, \apjl, 472, L107 

\bibitem[Belloni et al.(1997)]{bel97} Belloni, T., van der 
Klis, M., Lewin, W.~H.~G., et al. 1997, \aap, 322, 857

\bibitem[Belloni et al.(2002)]{bel02} 
Belloni, T., Psaltis, D., \& van der Klis, M.\ 2002, \apj, 572, 392 

\bibitem[Churazov et al.(2001)]{chu01} 
Churazov, E., Gilfanov, M., \& Revnivtsev, M.\ 2001, \mnras, 321, 759

\bibitem[Cui (1999)]{cui99} Cui, W., 1999, in High Energy Processes in Accreting Black Holes, J. Poutanen and R. Svensson eds., ASP Conf. Ser., 161 (San Francisco:ASP)

\bibitem[Cui et al.(1997a)]{cui97a} Cui, W., Heindl, W.~A., 
Rothschild, R.~E., et al. 1997, \apjl, 474, L57

\bibitem[Cui et al.(1997b)]{cui97b} Cui, W., Zhang, S.~N., Focke, W., \& Swank, J.~H.\ 1997, \apj, 484, 383 

\bibitem[Cui et al.(2002)]{cui02} Cui, W., Feng, 
Y., \& Ertmer, M.\ 2002, \apjl, 564, L77

\bibitem[di Matteo \& Psaltis(1999)]{MP99} di Matteo, T.~\& 
Psaltis, D.\ 1999, \apjl, 526, L101

\bibitem[Ebisawa et al.(1996)]{ebi96} Ebisawa, K., Ueda, Y., 
Inoue, H., Tanaka, Y., \& White, N.~E.\ 1996, \apj, 467, 419

\bibitem[Esin et al.(1998)]{esi98} Esin, A.~A., Narayan, R., 
Cui, W., Grove, J.~E., \& Zhang, S.\ 1998, \apj, 505, 854 

\bibitem[Gierli\'nski et al.(1997)]{gie97} Gierli\'nski, M., 
Zdziarski, A.~A., Done, C., Johnson, W.~N., Ebisawa, K., Ueda, Y., Haardt, 
F., \& Phlips, B.~F.\ 1997, \mnras, 288, 958 

\bibitem[Gierli{\'n}ski et al.(1999)]{gie99} Gierli{\'n}ski, M., Zdziarski, 
A.~A., Poutanen, J., et al. 1999, \mnras, 309, 496 

\bibitem[Gilfanov et al.(1999)]{gil99} 
Gilfanov, M., Churazov, E., \& Revnivtsev, M.\ 1999, \aap, 352, 182

\bibitem[Jahoda et al.(1996)]{jah96} Jahoda, K., Swank, J.~H., Giles, A.~B., et al. 1996, in EUV, X-Ray, and Gamma-Ray Instrumentation for Astronomy VII, O.H Siegmund ed., Proc. SPIE, 2808 (Bellingham, WA: SPIE), 59

\bibitem[Jernigan et al.(2000)]{jer00} Jernigan, 
J.~G., Klein, R.~I., \& Arons, J.\ 2000, \apj, 530, 875

\bibitem[Kato(1990)]{Kato90} Kato, S.\ 1990, \pasj, 42, 99

\bibitem[Kato(2001)]{kat01} Kato, S.\ 2001, \pasj, 53, 1

\bibitem[Klein-Wolt(2004)]{mkw04} Klein-Wolt, M. 2004, Ph.D. thesis, Anton Pannekoek Astronomical Institute and Center for High Energy Astrophysics, University of Amsterdam, Amsterdam

\bibitem[Miyamoto et al.(1992)]{miy92} Miyamoto, S., 
Kitamoto, S., Iga, S., Negoro, H., \& Terada, K.\ 1992, \apjl, 391, L21

\bibitem[Nowak(2000)]{now00} Nowak, M.~A.\ 2000, \mnras, 318, 
361 (N00)

\bibitem[Nowak et al.(1999)]{now99} Nowak, M.~A., Vaughan, 
B.~A., Wilms, J., Dove, J.~B., \& Begelman, M.~C.\ 1999, \apj, 510, 874

\bibitem[Okazaki et al.(1987)]{OKF87} Okazaki, 
A.~T., Kato, S., \& Fukue, J.\ 1987, \pasj, 39, 457

\bibitem[Pottschmidt et al.(2000)]{pot00} Pottschmidt, K., Wilms, J., Nowak, M.~A., et al.\ 2000, \aap, 357, L17

\bibitem[Pottschmidt et al.(2003)]{pot03} Pottschmidt, K., Wilms, J., Nowak, M.~A., et al.\ 2003, \aap, 407, 1039 (P03)

\bibitem[Poutanen(2001)]{pou01} Poutanen, J.\ 2001, Advances 
in Space Research, 28, 267

\bibitem[Poutanen et al.(1997)]{pou97} Poutanen, 
J., Krolik, J.~H., \& Ryde, F.\ 1997, \mnras, 292, L21

\bibitem[Psaltis \& Norman(2000)]{psa00} Psaltis, D. \& Norman, C. 2000, ArXiv Astrophysics e-prints, astro-ph/0001391

\bibitem[Psaltis et al.(1999)]{pbk99} 
Psaltis, D., Belloni, T., \& van der Klis, M.\ 1999, \apj, 520, 262

\bibitem[Reig et al.(2002)]{rei02} Reig, P., 
Papadakis, I., \& Kylafis, N.~D.\ 2002, \aap, 383, 202

\bibitem[Shapiro \& Lightman(1976)]{ShL76} Shapiro, S.~L.~\& 
Lightman, A.~P.\ 1976, \apj, 204, 555 

\bibitem[Shapiro et al.(1976)]{sha76} Shapiro, 
S.~L., Lightman, A.~P., \& Eardley, D.~M.\ 1976, \apj, 204, 187 

\bibitem[Stella \& Vietri(1998)]{SV98} Stella, L.~\& Vietri, 
M.\ 1998, \apjl, 492, L59 

\bibitem[Stella \& Vietri(1999)]{SV99} Stella, L.~\& Vietri, 
M.\ 1999, Physical Review Letters, 82, 17 

\bibitem[Stella, Vietri, \& Morsink(1999)]{SVM99} Stella, L., 
Vietri, M., \& Morsink, S.~M.\ 1999, \apjl, 524, L63 (SVM) 

\bibitem[van der Klis(1995)]{vdk95} van der Klis, M.\ 1995, in 
\textit{X-Ray Binaries}, W.~H.~G. Lewin, J. van Paradijs, \& E.~P.~J. van 
den Heuvel eds., Cambridge Univ. Press, Cambridge 

\bibitem[van der Klis(2004)]{vdk04} van der Klis, M.\ 2004, to appear in Compact stellar X-ray sources, Lewin \& van der Klis eds., Cambridge University Press, astro-ph/0410551

\bibitem[van Straaten(2004)]{vst04} van Straaten, S. 2004, Ph.D. thesis, Anton Pannekoek Astronomical Institute and Center for High Energy Astrophysics, University of Amsterdam, Amsterdam

\bibitem[van Straaten et al.(2002)]{vst02} van Straaten, S., van der Klis, M., di Salvo, T., \& Belloni, T.\ 2002, \apj, 568, 912 

\bibitem[Vikhlinin et al.(1994)]{vik94} 
Vikhlinin, A., Churazov, E., \& Gilfanov, M.\ 1994, \aap, 287, 73

\bibitem[Wagoner(1999)]{wag99} Wagoner, R.V.\ 1999, \physrep, 311, 259p

\bibitem[Wagoner et al.(2001)]{wag01} Wagoner, R.~V., 
Silbergleit, A.~S., \& Ortega-Rodr{\'{\i}}guez, M.\ 2001, \apjl, 559, L25

\bibitem[Wen et al.(2001)]{wen01} Wen, L., Cui, W., 
\& Bradt, H.~V.\ 2001, \apjl, 546, L105

\bibitem[Wijnands(2001)]{wij01} Wijnands, R.\ 2001, Advances 
in Space Research, 28, 469

\bibitem[Wijnands \& van der Klis(1999)]{wvdk99} Wijnands, 
R.~\& van der Klis, M.\ 1999, \apj, 514, 939

\bibitem[Zdziarski \& Gierli{\'n}ski(2004)]{zg04} Zdziarski, 
A.~A.~\& Gierli{\'n}ski, M.\ 2004, Prog. Theor. Phys. Suppl., 155, 99-119

\bibitem[Zdziarski et al.(2002)]{zdz02} 
Zdziarski, A.~A., Poutanen, J., Paciesas, W.~S., \& Wen, L.\ 2002, \apj, 
578, 357

\bibitem[Zdziarski et al.(2004)]{zdzprep} Zdziarski, A.~A. et al. 2004, in preparation

\bibitem[Zhang et al.(1995)]{zha95} Zhang, W., Jahoda, K., 
Swank, J.~H., Morgan, E.~H., \& Giles, A.~B.\ 1995, \apj, 449, 930

\bibitem[Zhang et al.(1996)]{zha96} Zhang, W., Morgan, E.~H., 
Jahoda, K., et al.\ 1996, \apjl, 469, L29

\bibitem[Zhang et al.(1997)]{ZCC97} Zhang, S.~N., Cui, 
W., \& Chen, W.\ 1997, \apjl, 482, L155 

\end{thebibliography}
\end{document}